\documentclass[titlepage,12pt]{article}
\usepackage{textcomp}
\usepackage{chetnew}
\usepackage{amssymb,amsmath,color,graphics,amscd,epsf,indentfirst,amsfonts}
\usepackage{mathtools}
\usepackage{enumerate}
\usepackage{epsfig}
\usepackage{slashed}

\usepackage{tikz}

\def\blfootnote{\xdef\@thefnmark{}\@footnotetext}

\long\def\symbolfootnote[#1]#2{\begingroup%
\def\thefootnote{\fnsymbol{footnote}}\footnote[#1]{#2}\endgroup}

\makeatletter
\renewcommand{\@dotsep}{4.5}
\makeatother

\def\be{\begin{equation}}
\def\ee{\end{equation}}

\makeatletter
\def\@seccntformat#1{\csname the#1\endcsname.\quad}
\makeatother

\def\clock{{\count0=\time
           \divide\count0 60
           \ifnum\count0<10 0\fi\the\count0
           \multiply\count0 -60 \advance\count0 \time
           :\ifnum\count0<10 0\fi \the\count0
         }}
\newcommand{\timestamp}{{\small\vbox{\hbox{\tt\jobname.tex}
\hbox{\the\day/\the\month/\the\year, \clock}}}}

\def\LL{{\cal L}}

\def\NN{{\cal N}}

\def\Tr{{\rm {Tr}}}

\def\p{{\partial}}
\def\time{{\tau}}

\def\beq{\begin{equation}}
\def\eeq{\end{equation}}
\newcommand{\bea}{\begin{eqnarray}}
\newcommand{\eea}{\end{eqnarray}}
\def\bal{\begin{align}}
\def\eal{\end{align}}

\newcommand{\drawsquare}[2]{\hbox{%
\rule{#2pt}{#1pt}\hskip-#2pt
\rule{#1pt}{#2pt}\hskip-#1pt
\rule[#1pt]{#1pt}{#2pt}}\rule[#1pt]{#2pt}{#2pt}\hskip-#2pt
\rule{#2pt}{#1pt}}

\newcommand{\Yfund}{\raisebox{-.5pt}{\drawsquare{6.5}{0.4}}}
\newcommand{\Yasymm}{\raisebox
{-3.5pt}{\drawsquare{6.5}{0.4}}\hskip-6.9pt%
                      \raisebox{3pt}{\drawsquare{6.5}{0.4}}%
                     }
\newcommand{\Ysymm}{\Yfund\hskip-0.4pt%
                     \Yfund}

\newcommand\T{\rule{0pt}{2.5ex}}

\newcommand\B{\rule[-1.7ex]{0pt}{0pt}}

\def\drawbox#1#2{\hrule height#2pt
         \hbox{\vrule width#2pt height#1pt \kern#1pt
               \vrule width#2pt}
               \hrule height#2pt}

\def\Asym#1#2{\vcenter{\vbox{\drawbox{#1}{#2}
               \kern-#2pt       
               \drawbox{#1}{#2}}}}

\def\bdot{\huge{\textbf{.}}}

\numberwithin{equation}{section}

\begin{document}
\begin{titlepage} 
\begin{flushright}
DCPT-17/35
\vskip -1cm
\end{flushright}
\vskip 4cm
\begin{center}
\font\titlerm=cmr10 scaled\magstep4
    \font\titlei=cmmi10 scaled\magstep4
    \font\titleis=cmmi7 scaled\magstep4
    \centerline{\LARGE \titlerm 
      Phases of QCD$_3$ from Non-SUSY Seiberg Duality}
      \vskip 0.3cm
    \centerline{\LARGE \titlerm  and Brane Dynamics}
    \vskip 0.3cm
\vskip 1cm
{Adi Armoni$^\star$ and Vasilis Niarchos$^\natural$}\\
\vskip 0.5cm
       {\it $^\star$Department of Physics, College of Science}\\
       {\it Swansea University, SA2 8PP, UK}\\
\medskip
{\it $^\natural$Department of Mathematical Sciences and Center for Particle Theory}\\
{\it Durham University, Durham, DH1 3LE, UK}\\
\medskip
\vskip 0.5cm
{$^\star$a.armoni@swansea.ac.uk, $^\natural$vasileios.niarchos@durham.ac.uk}\\

\end{center}
\vskip .5cm
\centerline{\bf Abstract}

\baselineskip 20pt
%

\vskip .5cm 
\noindent
We consider a non-supersymmetric USp Yang-Mills Chern-Simons gauge theory coupled to fundamental flavours. The theory is realised in type IIB string theory via an embedding in a Hanany-Witten brane configuration which includes an orientifold and {\it anti} branes. We argue that the theory admits a magnetic Seiberg dual. Using the magnetic dual we identify dynamics in field theory and brane physics that correspond to various phases, obtaining a better understanding of 3d bosonization and dynamical breaking of flavour symmetry in USp QCD$_3$ theory. In field theory both phases correspond to magnetic 'squark' condensation. In string theory they correspond to open string tachyon condensation and brane reconnection. We also discuss other phases where the magnetic 'squark' is massive. Finally, we briefly comment on the case of unitary gauge groups.

\vfill
\noindent
\end{titlepage}\vfill\eject

\hypersetup{pageanchor=true}

\setcounter{equation}{0}

\pagestyle{empty}
\small
\vspace*{-0.7cm}
{
\hypersetup{linkcolor=black}
\tableofcontents
}
\normalsize
\pagestyle{plain}
\setcounter{page}{1}

\section{Introduction and conclusions} 
\label{intro}

Quantum chromodynamics in three spacetime dimensions (QCD$_3$) is an interesting variant of the more familiar QCD in four dimensions. In a sense, the three dimensional theory is richer because three dimensions allow the addition of a topological Chern-Simons (CS) term which blocks the RG running of the Yang-Mills interaction towards strong coupling and alters the infra-red (IR) dynamics.

In the past decade or so there has been significant progress in understanding the dynamics of 3d supersymmetric gauge theories including the dynamics of a large class of  supersymmetric Chern-Simons theories coupled to different types of matter. Steps forward have also been achieved in the study of non-supersymmetric Yang-Mills-Chern-Simons (YM-CS) theories like QCD$_3$. For example, consider a CS theory coupled to $N_f$ Dirac fermions in the fundamental representation. A recent intriguing result is a conjectured duality between the following pairs of theories\cite{Aharony:2011jz,Giombi:2011kc,Aharony:2012nh,Radicevic:2015yla,Hsin:2016blu} 
\beq
\label{introaa}
SU(N_c)_{K+\frac{N_f}{2}}~\oplus~N_f~{\rm fermions} ~~\longleftrightarrow~~ U(K+\frac{N_f}{2})_{-N_c}~\oplus~N_f~{\rm scalars}
~.\eeq
A variant of this duality postulates that
\bea
\label{introaab}
U(N_c)_{K+\frac{N_f}{2},K+\frac{N_f}{2} \pm N_c}  \oplus N_f~{\rm fermions}
\longleftrightarrow
U(K+\frac{N_f}{2})_{-N_c,-N_c \mp (K+\frac{N_f}{2})} \oplus N_f~{\rm scalars}.
\eea

In these expressions, and in what follows, the notation $SU(N)_k$ quotes {\it bare} CS levels. These are written in terms of the shifted level $K$ defined as $k_{bare}-\frac{N_f}{2}$. Similarly, in the notation $U(N)_{k,\ell}$ the first level refers to the bare level of the $SU$ part and the second to the bare level of the $U(1)$ part.
Since both of these dualities relate a theory with fermions to a theory with bosons they are frequently referred to as bosonization (in close analogy with the more familiar bosonization in two dimensions). The scalars on the bosonic side have quartic interactions tuned to a Wilson-Fisher fixed point. The duality holds for $\frac{1}{2}N_f \leq K$.

Versions of this duality for $SO(N_c)$ and $USp(2N_c)$ gauge groups have also been formulated \cite{Aharony:2016jvv}. In this paper, we will mostly focus on results that are very closely related to the $USp(2N_c)$ version of the duality

\beq
\label{introab}
USp(2N_c)_{2K+N_f}~\oplus~2N_f~{\rm fermions} ~~\longleftrightarrow~~ USp(2K+N_f)_{-2N_c}~\oplus~2N_f~{\rm scalars}
~.\eeq
Similar to the $SU(N_c)$ case, the bosonic side requires that the scalars are at a $USp(2N_f)$ invariant Wilson-Fisher fixed point. This duality is also expected to be valid for $\frac{1}{2}N_f \leq K$. 
  
More recently, Komargodski and Seiberg (KS) argued for a scenario \cite{Komargodski:2017keh} that describes the infra-red dynamics of the fermionic theory when $N_f >2|K|$. According to that scenario there is a window, for $2 |K| < N_f < N_*$, where the theory exhibits a phase of flavor symmetry breaking: $U(N_f) \rightarrow U(N_f/2-K)\times U(N_f/2+K)$ in the unitary case and $USp(2N_f) \rightarrow USp(N_f-2K)\times USp(N_f+2K)$ in the symplectic case. $N_*$ is a critical number of flavors whose precise value is currently unknown. The evidence in favour of this scenario includes the matching of anomalies and consistency under several RG flows \cite{Komargodski:2017keh}. For $N_f >N_*$ it is conjectured \cite{Appelquist:1988sr,Appelquist:1989tc,Komargodski:2017keh} that the theory flows to some IR fixed point that does not exhibit flavor symmetry breaking. 

\subsection{Summary of results}

\subsection*{Benefits of an ultra-violet embedding} 

In this paper we provide new evidence in favour of the above scenario for QCD$_3$ (bosonization, symmetry breaking, CFT) by embedding the IR dynamics of QCD$_3$ in an ultra-violet (UV) YM-CS theory, which is a 3d cousin of a 4d non-supersymmetric orientifold QCD theory \cite{Armoni:2013ika}. The main part of the paper will focus on the case of $USp(2N_c)$ gauge group. In that case, the UV embedding involves a $USp(2N_c)$ YM-CS theory at bare level $2k$ coupled to a real scalar field in the 2-index symmetric representation (`scalar gaugino'), a Dirac fermion in the 2-index antisymmetric representation (`gaugino'), $2N_f$ scalars in the fundamental representation (`squarks') and $2N_f$ fermions in the fundamental representation (`quarks'). We argue that there are regimes when $k\neq 0$ where the quantum effects lift the scalars and the IR behaviour is dominated by standard QCD$_3$ physics. The level $K$ that appears in KS \cite{Komargodski:2017keh}, e.g.\ in eq.\ \eqref{introab}, is related to $k$ via the relation 
\beq
\label{introac}
K = | k+1-N_c | -\frac{N_f}{2}
~.
\eeq
The combination $k+1-N_c$ arises naturally when we integrate out the antisymmetric gaugino and will appear frequently in our formulae. It is therefore convenient to introduce the integer
\beq
\label{introad}
\kappa \equiv k+1-N_c 
~,\eeq
in terms of which $K= |\kappa|- \frac{N_f}{2}$.

The great advantage of the additional degrees of freedom in the orientifold QCD theory is that they allow us to formulate a non-supersymmetric version of Seiberg duality. For $N_f=0$ this duality reduces to level-rank duality \cite{Armoni:2014cia}. For general $N_f$ the dual `magnetic' theory is a $USp(2N_f+2\kappa)$ gauge theory whose details will be described in detail in section \ref{duality}. 4d versions of the duality were studied in \cite{Armoni:2013ika}. In this paper we will show that the finite $N_f$ duality passes the standard checks of consistency under RG flows, global symmetry and 't Hooft anomaly matching. Further favourable evidence is provided by string theory. We will return to some of the salient features of the string theory embedding in a moment.

The electric-magnetic duality operates when the rank of the dual gauge group is positive, namely when $N_f+k+1 > N_c$ (equivalently, when $N_f > - \kappa$). For values of $k,N_c,N_f$ outside this window there is no magnetic dual and one has to analyse separately the strong coupling dynamics of the electric theory. It is unclear what the infra-red physics of the theory are in this regime. In supersymmetric analogues, e.g.\ in 3d $\NN=2$ SQCD theories, the supersymmetric vacuum is lifted by non-perturbative effects when the dual rank becomes negative. When the dual rank vanishes the dual theory is a theory of free chiral multiplets. 

In this paper we will work exclusively in the regime where the dual magnetic description exists. The magnetic description will provide an illuminating perspective on the dynamics of the electric theory. In particular, we will find, under certain assumptions, that the magnetic theory leads to a rather natural universal description for all the phases of QCD$_3$ outlined above. We will show that when $N_f \leq 2|K|$ bosonization emerges naturally in the IR via magnetic squark condensation; a mechanism that reminds of monopole condensation and the dual Meissner effect in four dimensional physics. We will not be able to prove conclusively the existence of magnetic squark condensation, but we will provide evidence indicating that it is a viable possibility and that it leads to a suggestive consistent synthesis of known results. 

\begin{figure}[!t]
\begin{picture}(100,180)
\put(10,10){\includegraphics[width=6.5cm]{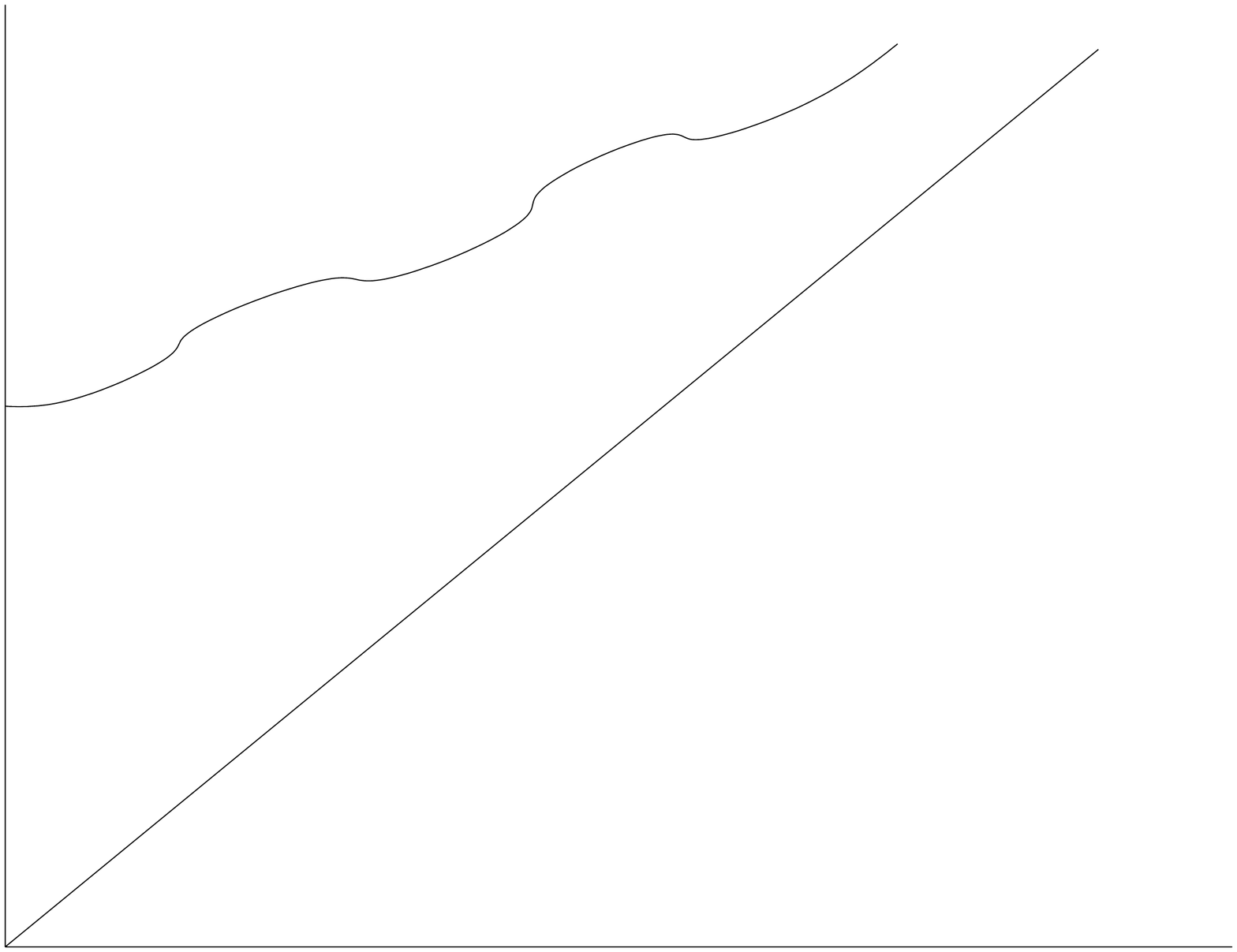}}
\put(5,165){$N_f$}
\put(140,155){$N_*$}
\put(180,155){$N_f = |\kappa|$}
\put(205,10){$|\kappa|$}
\put(70,-20){A: QCD$_3$ phases}
\put(150,50){I}
\put(40,70){II}
\put(50,130){III}
\end{picture}
\begin{picture}(100,150)
\put(140,0){\includegraphics[width=7.5cm]{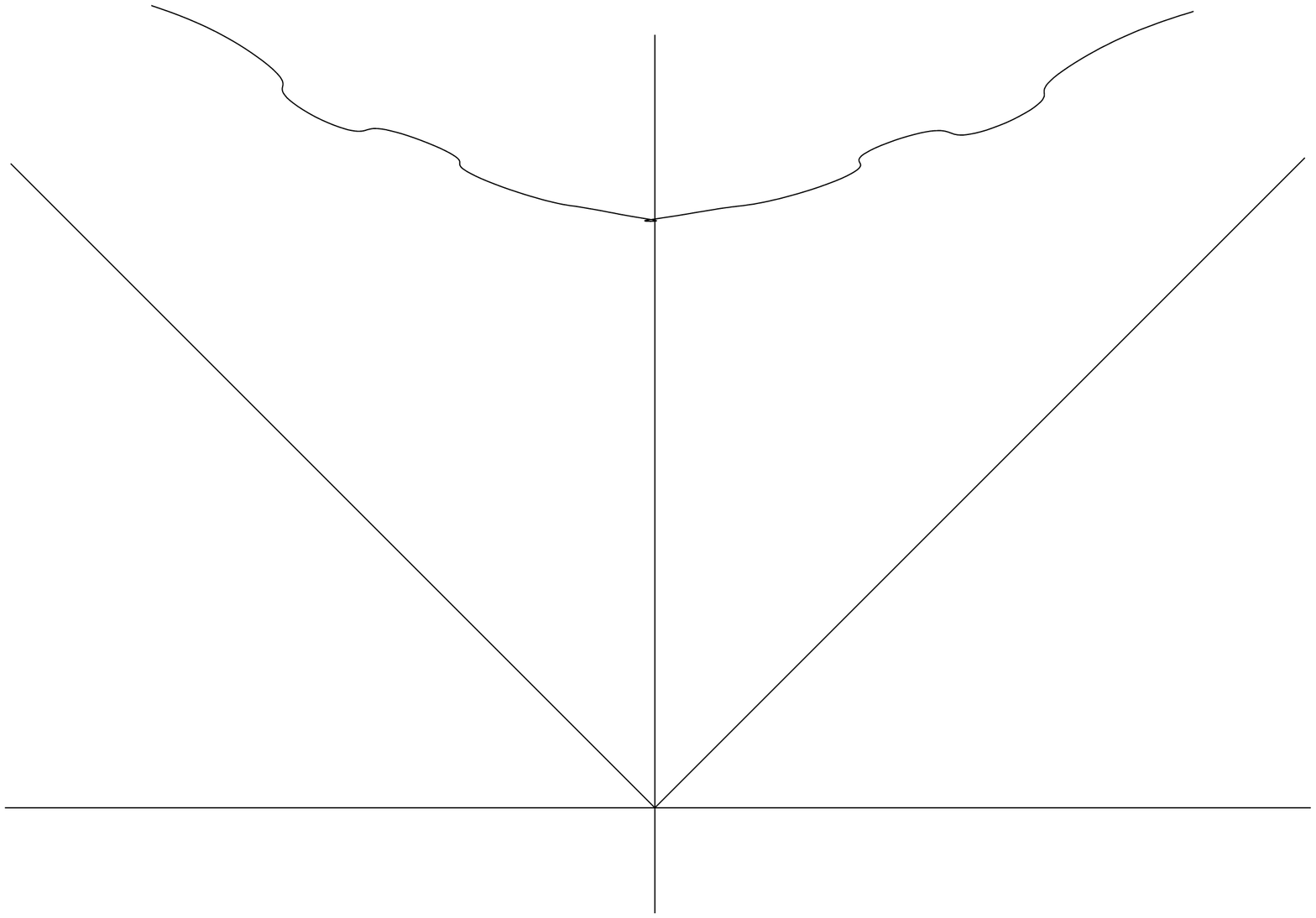}}
\put(240,155){$N_f$}
\put(335,155){$N_*$}
\put(350,135){$N_f = |\kappa|$}
\put(360,15){$\kappa$}
\put(180,-20){B: Orientifold QCD$_3$ phases}
\put(125,40){\small\it no Seiberg duality}
\put(320,50){I}
\put(280,90){II}
\put(200,90){II$'$}
\put(290,155){III}
\put(185,155){III}
\end{picture}
\vspace{1cm}
\caption{\footnotesize Diagram A represents the phases of QCD$_3$ anticipated by \cite{Komargodski:2017keh}. Region I is the phase of bosonization, region II the phase of symmetry breaking and region III a phase with an IR CFT description. The curve separating the regions II and III occurs at some $N_*$ which is potentially a function of both $\kappa$ and $N_c$. The wiggly features of this curve are not a statement about its actual shape, but rather a symbolic depiction of our ignorance about its precise form. A constraint on the shape of this curve was determined in Ref.\ \cite{Komargodski:2017keh}. Diagram B represents the phases of orientifold QCD$_3$ that arise naturally (under a certain set of assumptions) from the magnetic description of the theory. With the exception of the region $\kappa<-N_f$ where no Seiberg duality is available the remaining regions I, II/II$'$ and III, are expected to describe the same IR physics as the corresponding regions in diagram A.}
\label{fig_phases}
\end{figure}

This mechanism provides a new explanation of 3d bosonization. Unlike previous explanations based on deformations of supersymmetric 3d Seiberg dualities and mirror symmetry (see e.g.\ \cite{Jain:2013gza,Gur-Ari:2015pca,Kachru:2016rui,Kachru:2016aon} for beautiful work in these directions) in this mechanism bosonization arises dynamically in the infrared as a consequence of a non-supersymmetric version of Seiberg duality. These effects describe the part of the phase diagram denoted as I in the diagrams of Fig.\ \ref{fig_phases}.

Once magnetic squark condensation is assumed the phase of symmetry breaking for $N_f > 2 |K|$ follows naturally. There are two regions in the phase diagram of orientifold QCD$_3$, regions II and II$'$ in diagram B of Fig.\ \ref{fig_phases}, that describe in the IR the physics of QCD$_3$ for $N_f > 2|K|$.

Interestingly, the magnetic description in regions II and II$'$ is not identical. Consider first the situation in region II$'$ where $-N_f < \kappa <0$. In this case squark condensation leads to a complete Higgsing of the gauge group leaving behind a sigma model of Goldstone bosons for the dynamically broken symmetry. The magnetic theory provides an explicit description of how the sigma model arises.

Region II of the phase diagram of orientifold QCD$_3$ refers to the parameter regime $N_f > \kappa >0$. 

The dictionary \eqref{introac} implies that this region describes the same IR physics as QCD$_3$ with $N_f > 2|K|$. Naively, it looks like a natural continuation of the bosonization regime where squark condensation leads to a theory of $2N_f$ bosons. It is obvious, however, that this theory cannot be the same as the bosonic dual of region I, because bosonization is inconsistent for $N_f>2|K|$, which is also clear in the behaviour of the electric and magnetic orientifold QCD theories under mass deformations. This suggests that $N_f = \kappa$ is a critical point above which the magnetic theory reverts to the symmetry breaking phase. We propose a particular scenario for the mechanism behind this phase transition. This scenario is consistent with expectations about the massive deformations of the fermionic (electric) theory and predicts naturally a phase transition to a topological sector as the mass of the fermions is increased. A dual bosonic formulation appears at the transition point as anticipated by KS \cite{Komargodski:2017keh}. 

The symmetry pattern for sufficiently large $N_f$ persists as long as the magnetic squarks are tachyonic and condense. Competing effects in the one-loop mass normalisation of the squarks suggest the possibility that when $N_f$ is large enough, $N_f \geq N_*$ for some $N_*(k,N_c)$, the squarks become massive. Depending on the mass squared of the elementary magnetic meson fields, Seiberg duality suggests a specific description for the IR theory. We argue that the most likely scenario is one where the elementary magnetic mesons are tachyonic. The condensation of the elementary magnetic mesons leads to an IR theory of free fermions and light mesons, a scenario that agrees with the large-$N_f$ results in \cite{Appelquist:1988sr,Appelquist:1989tc}.

\subsection*{Lessons from string theory}

It is useful to consider a further UV embedding of the orientifold QCD theory into a non-supersymmetric brane configuration in string theory. The immediate benefits of this embedding are:
\begin{itemize}
\item[$(a)$] String theory is a concrete guide to the tree-level Lagrangian of the magnetic dual and provides additional evidence in favour of the non-supersymmetric Seiberg duality that is the center-point of this paper. 
\item[$(b)$] Non-trivial effects in field theory have a natural description in terms of brane physics (frequently as a geometric rearrangement of the brane configuration). For example, the magnetic squark condensation that drives the UV theory to bosonization in the IR has a natural interpretation in brane physics as {\it open string tachyon condensation and brane reconnection.}\footnote{A different embedding of 3d bosonization in string theory was recently proposed in ref.\cite{Jensen:2017xbs} which describes via holography effects in gauge theory in the large-$N$ limit.} 
\end{itemize}

In this paper we are interested in non-supersymmetric brane configurations in a Hanany-Witten brane setup in type IIB string theory along the lines of the Giveon-Kutasov analysis \cite{Giveon:2008zn}. In order to break supersymmetry we consider a combination of an $O3$ plane, {\it anti} D5 branes and {\it anti} D3 branes suspended between two 5-branes, one of which is a bound state of an NS5 brane and {\it anti} D5 branes.\footnote{A similar 4d brane configuration in type IIA string theory and a corresponding non-supersymmetric Seiberg duality was proposed in \cite{Armoni:2013ika}.} The mutual presence of anti-branes with an orientifold plane breaks the ${\cal N}=2$ supersymmetry completely. Supersymmetry is restored asymptotically when $N_c,N_f$ and $k$ are taken to infinity. This feature is useful. The large-$N$ regime is a technically convenient regime where supersymmetry is softly broken by $1/N$ effects.\footnote{In this paper the conjectured non-supersymmetric Seiberg duality is applied at finite-$N$ and the statements resulting from this application are also intended as statements for the finite-$N$ theory.}

The broken supersymmetry leads to non-trivial potentials between different components of the brane configuration. On the electric side the potentials are attractive and lead to a stable brane configuration. On the magnetic side, however, the potentials can be repulsive, leading to open string tachyon condensation and brane reconnection between the flavor and colour branes. The rich phase diagram that arises in this manner is literally a web of phases driven by the presence or absence of open string instabilities. The result translates in gauge theory either as bosonization or dynamical symmetry breaking, or a CFT.

\subsection{Outline of the paper}

The paper is organised as follows. In section \ref{duality} we present the orientifold field theories of interest, describe how they are embedded in suitable brane configurations in ten-dimensional type IIB string theory, and formulate the non-supersymmetric electric-magnetic duality that they are conjectured to obey. The evidence in favour of the duality is summarised in a separate subsection.

In section \ref{electric-magnetic} we discuss perturbative non-supersymmetric effects in both the electric and magnetic gauge theories. We discuss the expected behaviour of the electric theory in the IR and its relation to QCD$_3$. A detailed description of the IR physics from the magnetic theory point of view is relegated to the subsequent sections.

In section \ref{fullcond} we consider the parameter regime where $k+1-N_c \geq 0$. In that case of the magnetic squarks can lead to 'full color-flavor recombination' (color-flavor locking) on the magnetic side. This is translated to 3d bosonization of the electric side. The magnetic description does not provide only the correct matter content for bosonization, but also the requisite $USp(2N_f)$-invariant quartic interaction needed for the Wilson-Fischer fixed point. It is argued that this picture is consistent for $N_f \leq k+1-N_c$, but fails for larger values of $N_f$, where the most natural scenario, in accordance with expectations from QCD$_3$, is a scenario of a symmetry breaking phase.   

The other regime is when $-N_f \le k+1-N_c < 0$. In that regime the squark condensation Higgses the magnetic gauge group completely and leads to `partial color-flavor recombination'. Section \ref{partialcond} provides a detailed description of the effects that take place in this regime. On the electric field theory side these effects are translated as dynamical symmetry breaking. The Nambu-Goldstone bosons are identified as massless modes in the low-energy spectrum of the open string theory on the branes and as massless modes after squark condensation in the magnetic Lagrangian.

In section \ref{massivesquark} we discuss the possibility of a phase where the magnetic squark becomes massive. This leads to a phase with a CFT description in the IR in accordance with field theory expectations in QCD$_3$ \cite{Appelquist:1988sr,Appelquist:1989tc}. The magnetic dual provides a specific prediction for the IR CFT. This possibility requires the existence of a new critical number $N_*$ with the new phase being realised when $N_f \geq N_*$. 

In the above analysis it is always assumed that the bare CS level of the orientifold QCD theory $k$ is non-zero. When $k=0$ the IR physics is dominated by the YM interaction. The IR physics of this theory, which besides the effects of the fundamental fermions involves non-trivial dynamics from Dirac fermions in the 2-index anti-symmetric representation as well as additional scalar fields, is an interesting question that has not been explored in the past. This situation is discussed in section \ref{yangmills}. A recent discussion of CS theories with matter in the adjoint representation but no fundamentals can be found in \cite{Gomis:2017ixy}.

\subsection{Open problems}

In this paper we focus on the case of (orientifold) QCD$_3$ theories with symplectic gauge group. This choice is dictated by the fact that this case is the most straightforward one from the viewpoint of the string theory construction that underlies part of this work. Since the analysis of unitary groups with even rank, $U(2N)$, shares many similarities with the analysis of the $USp(2N)$ case (and is related to it by a non-perturbative planar equivalence), it is rather natural to put forward a corresponding picture for $U(2N)$ (orientifold) QCD$_3$. Preliminary comments in this direction are summarised in appendix \ref{unitary}.

The formulation of non-supersymmetric Seiberg duality for orientifold QCD$_3$ theories with general unitary gauge groups and orthogonal groups remains an open problem. We hope to return to these cases in a future publication.

\subsection{Summary of conventions}
\label{conventions}

In what follows when we mention the level of a UV gauge theory we will always refer, unless otherwise stated, to the bare level $k$ (or $2k$ in the $USp$ gauge theories). $k$ is always an integer ---it is the same integer (or related to the integer) that appears in the 5-brane bound states in our brane configurations. This convention is to be contrasted with other notation in the literature, e.g.\ \cite{Komargodski:2017keh} that is closely related to our discussion, where the quoted level is $K=k_{bare}+k_{quantum}$ with $k_{quantum}=- \frac{N_f}{2}$. 

When we integrate out massive particles the CS level shifts. In the YM-CS regularisation (which we follow) only the fermions contribute to the shift. We will use conventions natural in string theory where the integration of a single Dirac fermion with positive mass $m$ leads to the IR level
\beq
\label{introcaa}
k_{IR} = k+1
~.
\eeq
A list of possible mass deformations, and their string theory interpretation, is summarised in appendix \ref{RG}.

Another useful fact in our discussion is the fact that when we integrate out a gaugino (in the 2-index antisymmetric representation) in the $USp(2N)$ theory we obtain the CS level shift
\beq
\label{introcb}
k_{IR} = k - {\rm sgn}(k) (N-1)
~.\eeq

\section{Non-supersymmetric $USp$ Seiberg duality}
\label{duality}

\begin{figure}[!t]
\centerline{\includegraphics[width=12cm]{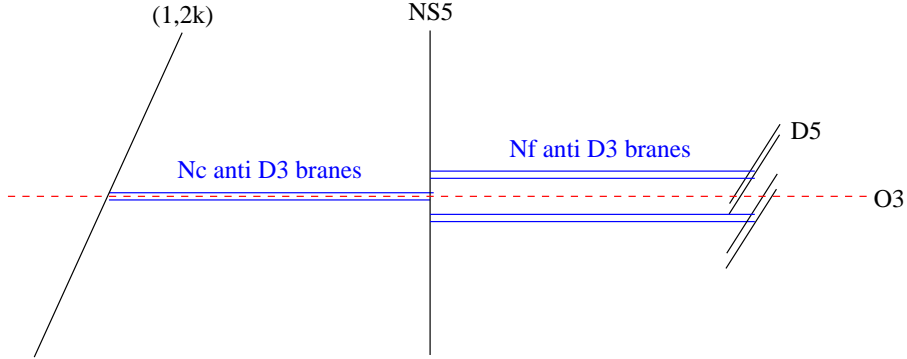}}
\caption{\footnotesize The electric theory. It is a non-supersymmetric $USp(2N_c)$ Yang-Mills theory with a level $2k$ Chern-Simons term and $2N_f$ flavours.}
\label{fig_electric}
\end{figure}

The gauge theories of interest can be phrased independently of string theory. Nevertheless, since string theory provides a convenient organising principle for the non-supersymmetric duality of interest we will present in parallel the gauge theories in question and their string theory embedding. We consider the $USp$ YM-CS theories that describe the IR physics on the Hanany-Witten brane configurations depicted in Figs.\ \ref{fig_electric}, \ref{fig_magnetic}. 

\subsection{Electric theory}

Let us start with a description of the electric theory. The brane configuration that engineers the theory consists of $N_c$ anti D3 branes suspended between an NS5 brane and a tilted $(1,2k)$ fivebrane. In addition there is an orientifold $O3$. The orientifold plane changes from $O3^+$ to $O3^-$ when it crosses the fivebrane. This is identical to the brane configuration of \cite{Armoni:2014cia}. In addition, we add $N_f$ anti D3 branes which are attached to the right of the NS5 brane and end on D5 branes. The resulting brane configuration is depicted in Fig. \ref{fig_electric}. It consists of

\begin{itemize}
\item an NS5 brane along the $012345$ directions,

\item a tilted $(1,2k)$ fivebrane along $012(37)89$ directions. The latter fivebrane is a bound state of an NS5 brane and $2k$ {\it anti}-D5 branes, it is tilted in the $(37)$ plane by an angle $\theta$ such that $\tan\left( \theta\right)=-2g_sk$. With this choice of angle, we would preserve ${\cal N}=2$ supersymmetry, had we not had an orientifold. The minus sign reflects the presence of anti-branes in the construction.
\item an $O3$ plane along the $0126$ directions, which is $O3^-$ at $x^6=\pm \infty$,
\item $N_c$ (color) {\it anti} D3 branes (and their mirrors) along the $012|6|$ directions ($|6|$ denotes that the branes have a finite extent in the 6-direction),
\item $N_f$ (flavor) {\it anti} D3 branes (and their mirrors) along the $012|6|$ directions. The flavor branes end on anti D5 branes, which are oriented along the $012789$ directions.
\end{itemize}

The matter content of the theory is similar to the matter content of the supersymmetric theory, except that the fields transform differently with respect to the gauge group, due to the presence of the orientifold. The full matter content of the electric theory is given in table \ref{tableelectric} below. Note that the gauge fields $A_\mu$ and the scalar gaugino $\sigma$ transform in the two-index symmetric (adjoint) representation of the gauge group, while the Dirac gaugino $\lambda$ transforms in the two-index antisymmetric representation. The complex scalars $\Phi$ (squarks) and the Dirac fermions $\Psi$ (quarks) are both in a fundamental pseudo-real representation of $USp(2N_c) \times SO(2N_f)$. Note that although the global symmetry on the brane is $SO(2N_f)$, the global symmetry of the low energy QCD theory is $USp(2N_f)$. This is due to irrelevant interactions inherent to the string realisation of the theory.

\begin{table}[!t]
\begin{center}
\begin{tabular}{|c|c|c|}
\hline
\multicolumn{3}{|c|} {Electric Theory} \\ 
\hline \hline
 & $Sp(2N_c)$ & $SO(2N_f)$  \\
\hline
 $A_\mu$ & \Ysymm & \bdot \\
& $N_c(2N_c+1)\B$ & \\
\hline
 $\sigma $ & \Ysymm & \bdot \\
& $N_c(2N_c+1)\B$ & \\
\hline
$\lambda$ & $\T\Yasymm\B$   &  \bdot \\
& $N_c(2N_c-1)\B$ & \\
                   \hline
$\Phi$   & $\Yfund$ & $\Yfund$ \\
&   $2N_c$  &   $2N_f$ \\ 
\hline
$\Psi$ & $\Yfund$ & $\Yfund$ \\
&   $2N_c$  &   $2N_f$  \\                   
\hline
\end{tabular}
\caption{\footnotesize The matter content of the electric theory. The $SO(2N_f)$ global symmetry of the brane construction enhances to $USp(2N_f)$ in the IR field theory.}
\label{tableelectric}
\end{center}
\end{table}

The classical Lagrangian of the electric theory is 
\beq
\label{eLagab}
\LL = \LL_{gauge} + \LL_{matter}
~.\eeq
The gauge part is Yang-Mills-Chern-Simons theory at level $2k$
\beq
\label{eLagac}
\LL_{gauge}= \LL_{YM} + \LL_{CS}
~,
\eeq
\beq
\label{eLagad}
\LL_{YM} = \frac{1}{g^2} \Tr \bigg[ -\frac{1}{4} \left(F_{\mu\nu} \right )^2 + \frac{1}{2} \left( D_\mu \sigma \right)^2 + i \bar\lambda \slashed{D} \lambda - i \bar \lambda \left[ \sigma, \lambda \right] + \frac{1}{2} D^2 \bigg]
~,
\eeq
\beq
\label{eLagae}
\LL_{CS} = \frac{k}{4\pi} \Tr \bigg[ \varepsilon^{\mu\nu\rho} \left( A_\mu \p_\nu A_\rho - \frac{2i}{3} A_\mu A_\nu A_\rho \right) + 2 D\sigma - 2 \bar \lambda \lambda \bigg]
~.\eeq
$D_\mu$ is the standard gauge covariant derivative and the contraction of spinor indices has been kept implicit. 
For the matter part
\bea
\label{eLagaf}
\LL_{matter} &=& D_\mu \bar\Phi^{ai} D^\mu \Phi_{ai} + i \bar\Psi^{ai} \slashed{D} \Psi_{ai} - \bar\Phi^{ai} (\sigma^2)_a^{~b} \Phi_{bi} + \bar\Phi^{ai} D_a^{~b} \Phi_{bi} 
\nonumber\\
&&- \bar\Psi^{ai} \sigma_a^{~b} \Psi_{ai} + i \bar\Phi^{ai} \bar\lambda_a^{~b} \Psi_{bi} - i \bar\Psi^{ai} \lambda_a^{~b} \Phi_{bi}
~,
\eea
where $a,b,\ldots = 1,\ldots, 2N_c$ are color indices and $i,j,\ldots = 1,\ldots, 2 N_f$ are flavor indices. The pseudo-reality condition imposes on bosons and fermions
\beq
\label{pseudoreal}
\bar \Phi^{bj}= \Phi_{ai}\Omega^{ab} \tilde \Omega^{ij}
~,\eeq
with $\Omega$, $\tilde \Omega$ the $USp(2N_c)$, $USp(2N_f)$ symplectic tensors.

\subsection{Magnetic theory}
\label{magnetictheory}

\begin{figure}[!t]
\centerline{\includegraphics[width=12cm]{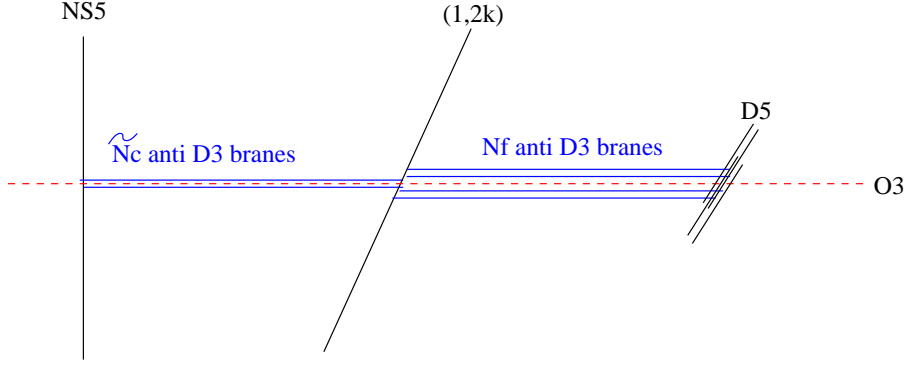}}
\caption{\footnotesize The magnetic theory. It is a non-supersymmetric $USp(2N_f+2k-2N_c+2)$ Yang-Mills theory with a level $-2k$ Chern-Simons term.}
\label{fig_magnetic}
\end{figure}

We define the proposed magnetic dual as the theory that describes the infra-red dynamics of the brane configuration in Fig.\ \ref{fig_magnetic}, which results from that in Fig.\ \ref{fig_electric} by swapping the NS5 and $(1,2k)$ 5-branes across the $x^6$ direction. In the presence of the orientifold $k+1$ additional D3 branes are created during this swap. As a result, we find $$\tilde N_c = N_f + k +1 -N_c =N_f + \kappa$$ D3 branes suspended between the NS5 and $(1,2k)$ 5-branes.

The dual theory is a $USp(2 \tilde N_c)$ gauge theory at bare CS level $-2k$. It exists when $\tilde N_c \geq 0$. Note that classically the flavor anti D3 branes can slide freely between the tilted fivebrane bound state and the $N_f$ D5 branes (which share the common directions (89))\footnote{As we discuss later, the moduli associated with motion in the (89) plane are lifted by quantum effects. Hence, they are pseudomoduli.} leading to corresponding elementary meson and mesino degrees of freedom. The meson $M$ transforms in the 2-index antisymmetric of $SO(2N_f)$ and the mesino $\chi$ in the 2-index symmetric representation of $SO(2N_f)$. The full matter content of the magnetic theory is summarised in table \ref{tablemagnetic}.

\begin{table}[!t]
\begin{center}
\begin{tabular}{|c|c|c|}
\hline
\multicolumn{3}{|c|} {Magnetic Theory} \\ 
\hline \hline
 & $\T Sp(2\tilde N_c)$ & $SO(2N_f)$  \\
\hline
 $a_\mu$ & \Ysymm & \bdot \\
& $\tilde N_c(2\tilde N_c+1)\B$ & \\
\hline
 $s$ & \Ysymm & \bdot \\
& $\tilde N_c(2\tilde N_c+1)\B$ & \\
\hline
$l$ & $\T\Yasymm\B$   &  \bdot  \\
& $\tilde N_c(2\tilde N_c-1)\B$ & \\
                   \hline
$\phi$   & $\Yfund$ & ${\Yfund}$ \\
&   $2\tilde N_c$  &   $2{N_f}$ \\
\hline
$\psi$ & $\Yfund$ & ${\Yfund}$ \\
&   $2\tilde N_c$  &   $2{N_f}$  \\
         \hline
 $M$ & \bdot & \Yasymm \\
& & $N_f(2N_f-1)$  \\
\hline
$\chi$ & \bdot & $\T\Ysymm\B$  \\
& & $N_f(2N_f+1)$ \\         
\hline
\end{tabular}
\caption{\footnotesize The matter content of the magnetic theory. The dual rank is expressed in terms of $\tilde N_c=N_f+k+1-N_c$.}
\label{tablemagnetic}
\end{center}
\end{table}

Let us discuss the classical Lagrangian of the magnetic theory. Besides the gauge interactions $\LL_{gauge}=\LL_{YM} + \LL_{CS}$ for gauge group $USp(2\tilde N_c)$
\beq
\label{mLagaa}
\LL_{YM} = \frac{1}{g^2} \Tr \bigg[ -\frac{1}{4} \left(f_{\mu\nu} \right )^2 + \frac{1}{2} \left( D_\mu s \right)^2 + i \bar l \slashed{D} l - i \bar l \left[ s, l \right] + \frac{1}{2} D^2 \bigg]
~,
\eeq
\beq
\label{mLagab}
\LL_{CS} = -\frac{k}{4\pi} \Tr \bigg[ \varepsilon^{\mu\nu\rho} \left( a_\mu \p_\nu a_\rho - \frac{2i}{3} a_\mu a_\nu a_\rho \right) + 2 Ds - 2 \bar l l \bigg]
\eeq
a part of the Lagrangian includes the same terms
\bea
\label{mLagac}
\LL_{matter,1} &=& D_\mu \bar\phi^{ai} D^\mu \phi_{ai} + i \bar\psi^{ai} \slashed{D} \psi_{ai} - \bar\phi^{ai} (s^2)_a^{~b} \phi_{bi} + \bar\phi^{ai} D_a^{~b} \phi_{bi} 
\nonumber\\
&&- \bar\psi^{ai} s_a^{~b} \psi_{ai} + i \bar\phi^{ai} \bar\lambda_a^{~b} \psi_{bi} - i \bar\psi^{ai} \lambda_a^{~b} \phi_{bi}
\eea
for the magnetic squarks $\phi^i$, and quarks $\psi^i$ as the electric theory. 
In addition, it includes kinetic terms for the mesons $M^{[ij]}$ and mesinos $\chi^{(ij)}$  
\beq
\label{mLagad}
\LL_{matter,2} = \p_\mu \bar M_{ij} \p^\mu M^{ij} + i \bar \chi_{ij} \slashed{\p} \chi^{ij}
\eeq
and a set of interactions that constitute the non-supersymmetric orientifold version of the supersymmetric cubic superpotential between mesons/mesinos and squarks/quarks.

The complete matter Lagrangian is
\bea
\label{mLagae}
\LL_{matter} &=& D_\mu \bar\phi^{ai} D^\mu \phi_{ai} + i \bar\psi^{ai} \slashed{D} \psi_{ai} - \bar\phi^{ai} (s^2)_a^{~b} \phi_{bi} + \bar\phi^{ai} D_a^{~b} \phi_{bi} 
\nonumber\\
&&- \bar\psi^{ai} s_a^{~b} \psi_{ai} + i \bar\phi^{ai} \bar\lambda_a^{~b} \psi_{bi} - i \bar\psi^{ai} \lambda_a^{~b} \phi_{bi} 
+  \p_\mu \bar M_{ij} \p^\mu M^{ij} + i \bar \chi_{ij} \slashed{\p} \chi^{ij} 
\nonumber\\
&& 
-\frac{1}{4}y^2 \phi_{ai} \phi^a_{~j} \bar\phi_b^{~i} \bar \phi^{bj} 
- y^2\bar M_{ik}M^{ij} \bar\phi_a^{~k} \phi^a_{~j} - 2y \psi_{ai}\chi^{ij} \phi_{~j}^{a} -y \psi_{ai} M^{ij} \psi_{~j}^{a}
\eea
and the total Lagrangian 
\beq
\label{mLagaf}
\LL = \LL_{gauge}+ \LL_{matter}
~.\eeq
$y$ is a coupling with mass dimension $1/2$ that appears in front of the cubic superpotential interactions in the supersymmetric version of the theory. In the non-supersymmetric theory at hand the RG flow will not respect the relations between the couplings that appear in front of the terms in the last line of \eqref{mLagae}. The expression \eqref{mLagae} is written here only as a specific bare Lagrangian that follows from its supersymmetric ancestor by appropriately modifying the representations of certain fields.

\subsection{Evidence for duality}

We claim that the above electric and the magnetic theories form a Seiberg dual. Apart from the argument in string theory based on swapping fivebranes, in appendix \ref{appEvidence} we provide standard evidence in the form of 't Hooft anomaly matching as well as a duality after deformations and RG flows.

Note that at large $N_c, N_f$ and $k$ the two theories become supersymmetric. The reason is that in this limit there is no difference between the two-index symmetric and the two-index antisymmetric representations. In the brane picture this statement translates to the fact that the orientifold (the M\"{o}bius amplitude) is a $1/N$ effect. Therefore, in the large $N$ limit the electric and magnetic theories are dual to each other as a supersymmetric pair similar to the one analysed in \cite{Giveon:2008zn}. 

There is another argument, from string theory, in favour of the duality, due to Sugimoto\footnote{Private discussions with AA about 4d non-supersymmetric Seiberg duality, IPMU, 2013.}. This argument does not involve fivebrane swapping. It relies on the SUSY duality and, under certain assumptions, leads to the non-SUSY duality. The idea is that the same operation (adding $N_f+k$ antibranes) on a pair of dual supersymmetric theories leads to a duality between non-supersymmetric electric and magnetic theories. The electric SUSY theory becomes the magnetic non-SUSY theory and the magnetic SUSY theory becomes the electric non-SUSY theory.

More explicitly, start from a SUSY Giveon-Kutasov brane configuration that realises a USp electric-magnetic pair \cite{Giveon:2008zn}. Consider the electric side with $N_c-1$ color branes and the magnetic side with $N_f+k-N_c$ colour branes. Both sides contain $N_f$ flavour branes. Add $N_f+k$ {\it infinite} anti-D3 branes on top of the SUSY electric theory and $N_f+k$ {\it infinite} anti-D3 branes on top of the SUSY magnetic theory. In both sides of the duality the $N_f+k$ antibranes extend over the flavour branes, the colour branes and beyond them where there are no D3 branes.

In the original electric side of the duality we obtain $N_f+k-N_c+1$ anti D3 branes as colour branes. Let us split the $N_f+k$ flavour segment of the anti branes into $N_f$ and $k$ anti branes. The $N_f$ anti branes annihilate the $N_f$ branes and the $k$ anti branes make the fivebrane bound state an NS5 brane. Similarly, on the ``other side'' of the brane configuration the $N_f+k$ antibranes become $N_f$ flavour antibranes and a tilted fivebrane bound state. The result is the magnetic $USp(2(N_f+k-N_c+1))$ non-SUSY theory.

In a similar way, adding $N_f+k$ antibranes to the colour part of the SUSY magnetic theory with $N_f+k-N_c$ colour brane results in $N_c$ antibranes. The annihilation of branes in the other segments of the configuration lead to the $USp(2N_c)$ non-SUSY electric theory.

We obtained the non-supersymmetric duality between the electric $USp(2N_c)$ theory and the magnetic $USp(2(N_f+k-N_c+1))$ theory from the SUSY pair. It will be interesting to repeat this exercise in a pure field theory language.

\section{Perturbative dynamics of the electric and magnetic theories}
\label{electric-magnetic}

Given that the electric and magnetic theories are non-supersymmetric we anticipate potentials for the various scalars.\footnote{Potentials, e.g.\ mass terms, are also generated for the fermions. These are typically subleading as a function of the UV cutoff. In this section we focus on the scalar potentials, which, by definition, can affect the stability of the theory.} The potentials are due to non-planar effects ($1/N$ effects) and they vanish in the large-$N$ limit where both the electric and magnetic theories become a dual pair with ${\cal N}=2$ supersymmetry. In some cases the field theory potentials have an interpretation as potentials between branes in the brane configuration. The potentials depend on the UV cut-off of the theory. Within field theory we can remove this dependence by renormalisation. The embedding in string theory, however, provides a natural UV cut-off, $\Lambda ^2= {1\over \alpha '}$, and a physical meaning to the potentials.

In all cases the effects that we consider are due to the difference between the  representations of bosons and fermions. Similar dynamics and considerations were involved in a proposal of a non-supersymmetric S-duality \cite{Sugimoto:2012rt}.

Consider a scalar propagator: in perturbation theory the difference between a bosonic loop and a fermionic loop will produce either a massive scalar or a tachyonic scalar, depending on whether there are more bosonic degrees of freedom or more fermionic degrees of freedom, as depicted schematically in Fig.\ \ref{mass} below.

\begin{figure}[!ht]
\centerline{\includegraphics[width=9cm]{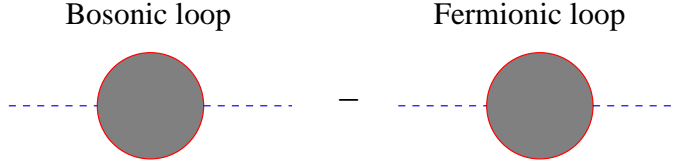}}
\caption{\footnotesize Perturbative contributions to a scalar mass.} 
\label{mass}
\end{figure}

If a scalar is massive it will not acquire a vev and the perturbative expansion near the origin (zero vev) is stable. If a scalar is tachyonic we anticipate a new minimum where a vev is acquired and symmetries (global or local) may be broken.

\subsection{Electric theory}

The electric theory contains two scalars, the squark $\Phi $ and the scalar gaugino $\sigma $.

The scalar gaugino couples to itself, to the gauge boson and to the fermionic gaugino $\lambda $. The generated mass due to the bosonic and fermionic loops is

\beq
M^2 _{\sigma} = g^2 (2N_c+2) \int ^\Lambda {d^3 k \over (2\pi)^3} {1\over k^2} -   g^2 (2N_c-2) \int ^\Lambda {d^3 k \over (2\pi)^3} {1\over k^2} \sim g^2 \Lambda \,
\eeq
namely $M^2 _{\sigma} >0$. This suggests aht the point $\langle \sigma \rangle =0$ is stable, and moreover, that the field $\sigma$ decouples from the low energy dynamics.

The potential $V(\sigma) \sim M^2\sigma ^2$ has an interpretation in the brane picture: the $N_c$ colour branes are attracted to the orientifold plane. The brane configuration is hence stable (non-tachyonic).

A similar analysis can be performed for the squark $\Phi$. The squark couples to the ``gauge multiplet'', namely to $A_{\mu}, \sigma, \lambda$. Since there are more bosonic than fermionic degrees of freedom perturbation theory suggests that $M^2 _{\Phi} >0$. The squark therefore decouples from the low energy physics.

As a result, the low energy field theory contains a $USp(2N_c)$ gauge field $A_{\mu}$, a gaugino $\lambda$ and $2N_f$ quarks $\Psi$.

When $k \ne 0$ there are Chern-Simons terms. The Chern-Simons terms provide a mass $M _{\rm CS}=g^2 k$ to the gauge field and to the gaugino. We therefore anticipate that the IR dynamics will be dominated by the topological Chern-Simons theory coupled to the quarks.\footnote{Integrating out the gaugino to obtain QCD$_3$ in the IR is most straightforward in the semiclassical regime $k\gg 1$. Away from this regime the relation with QCD$_3$ in the IR is harder to establish, but the expectation is that the IR phases are the same even for finite $k$ unless something drastic happens in the infrared behavior of OQCD$_3$. The breakdown of the magnetic description outside the window of Seiberg duality could be such an effect. Inside the window of Seiberg duality with $k\neq 0$ there are currently no indications of drastic effects that would lead to significant deviations from QCD$_3$ dynamics.} In that sense for $k \ne 0$ the electric theory is very similar to QCD$_3$.

The case $k=0$ is special, as there is no Chern-Simons term. The IR theory involves the strong coupling dynamics of the Yang-Mills interaction between the gauge field, the gaugino and the quarks.

\subsection{Magnetic theory}

Similarly to the electric theory, the scalar gaugino of the magnetic theory acquires a mass and decouples. The color branes are therefore attracted to the orientifold plane.

The dynamics of the squarks and the mesons is more complicated. Let us focus on the squarks.

At the one-loop level there are effects due to the coupling with the gauge multiplet and effects due to the coupling with the meson multiplet. Let us denote the magnetic gauge coupling by $g_m$ and the coupling to the meson multiplet by $y$.

There are more bosonic than fermionic degrees of freedom in the gauge multiplet and more fermionic than bosonic degrees of freedom in the meson multiplet. As a result,

\beq
M^2 _{\phi} \sim (-y^2 +g^2 _m) \Lambda \, . \label{squarkmass}
\eeq
At the one-loop level the two effects compete and the squark may become massive or tachyonic. We will consider both possibilities and identify the various phases associated with each one of them. We note that at large-$k$ the gauge field becomes very massive and decouples, therefore we anticipate that in this limit the dynamics is dominated by a tachyonic squark.

The magnetic theory includes a coupling of the form
\beq
y^2\bar M_{ik}M^{ij} \bar\phi_a^{~k} \phi^a_{~j} \,.
\eeq
 If the meson field acquires a vev of the form $ \langle  \bar M_{ik}M^{ij} \rangle  = v^2 \delta _k ^j$ the squark field becomes massive. If the squark field acquires a vev of the form $\langle \phi^a_{~j} \rangle = v\delta ^a_{~j} $, and if flavor symmetry is not broken (bosonized phase), the mesons become massive. We propose as the most likely scenario that in all phases
\beq
\label{mmrelation}
M^2 _ \phi \times M^2 _M <0 \, .
\eeq

In the following sections we will discuss several phases of the magnetic theory, their realisation in the brane picture and their relation to the electric theory.   

\section{Phases of orientifold QCD$_3$ and QCD$_3$}

\subsection{Bosonization}
\label{fullcond}

We focus on the magnetic description of the orientifold QCD$_3$ theory. The first region of parameter space under consideration is the region where $\kappa \equiv k+1 -N_c \geq N_f$, namely region I in diagram B of Fig.\ \ref{fig_phases}. The rank of the magnetic gauge group, $\tilde N_c = N_f + \kappa$ is automatically positive in this regime. The one-loop computation of the previous section suggests that the $2N_f$ squarks are tachyonic, at least for sufficiently small $N_f$ at fixed $\kappa$. Henceforth, we will operate under this assumption for the whole region I in diagram B of Fig.\ \ref{fig_phases}.

We assume that the magnetic squarks condense. Let us examine first what happens from the viewpoint of the brane configuration that describes the magnetic theory, Fig.\ \ref{fig_magnetic}. The magnetic squarks are low-lying modes of the open strings that stretch between the $N_f+\kappa$ color D3s to the $N_f$ flavor D3s (and their images). In the language of string theory condensation of the squarks is open string tachyon condensation through a process that reconnects the $2N_f$ color D3s with $2N_f$ flavor D3s. 

\begin{figure}[!t]
\centerline{\includegraphics[width=12cm]{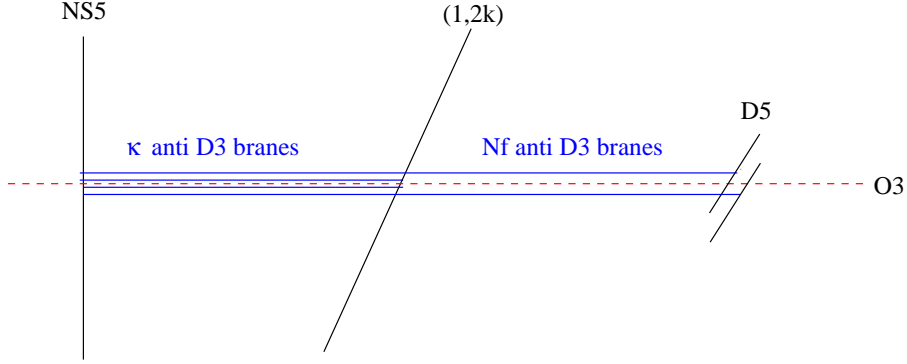}}
\caption{\footnotesize A realisation of the bosonized phase in the magnetic theory. $N_f$ flavor branes reconnect with color branes. The gauge group is $USp(2\kappa)$.}
\label{fig-bosonized}
\end{figure}

After reconnection $2N_f$ D3 branes are stretching between the NS5 brane and the $2N_f$ D5 branes, and $2\kappa$ color D3 branes are stretching between the NS5 brane and the $(1,2k)$ fivebrane, see Fig.\ref{fig-bosonized}.

Since the $2N_f$ reconnected D3 branes have no common directions with the fivebranes on which they end their low energy spectrum does not have a  gauge field, gaugino, mesons or mesinos. The only low lying modes arise from the strings on the $2\kappa$ color D3s and the strings stretching between the color D3s and the reconnected D3s. The theory on the color D3s is a $USp(2\kappa)$ gauge theory at level $-2k$ with a massive 2-index antisymmetric gaugino. The mass of the gaugino remains proportional to the level $2k$. At energies below this mass the gaugino can be integrated out. In addition, the field theory analysis below shows that the modes of the strings stretching between the color and reconnected flavor D3s are $2N_f$ scalars in the fundamental representation of the $USp(2N_f)$ gauge group. The reconnection preserves the original $SO(2N_f)$ global symmetry which enhances to $USp(2N_f)$ in the low energy theory.

In this manner brane reconnection implies that the IR physics is dominated by a $USp(2N_f)$ CS theory coupled to $2N_f$ bosons. Deformations of the brane setup that induce a non-vanishing mass to the quarks in the electric description show that the magnetic theory has the right interactions to be the bosonic theory appearing in bosonization. The deformation 2 in Fig.\ \ref{massFig} leaves the level unchanged and Higgses the dual gauge group. The deformation 4 shifts the level but does not Higgs the dual gauge group.

Let us take a closer look at the features of the magnetic field theory. In terms of the low-energy magnetic theory of section \ref{magnetictheory} it is easy to see that the color-flavor locking vev of the magnetic squarks Higgses the color gauge group from 
\beq
\label{bosonaa}
USp(2N_f + 2\kappa) ~ \to ~ USp(2\kappa)
\eeq
and leaves behind the gauginos in the 2-index antisymmetric of the Higgsed gauge group and $2N_f$ fundamental squarks. The $2N_f$ magnetic quarks become massive in the presence of the squark vev because of the Yukawa coupling between the squarks, the gauginos and the quarks. Similarly, both the elementary mesons and mesinos obtain masses and decouple at low energies.  

The resulting low energy theory has global symmetry $USp(2N_f)$. After integrating out the massive gaugino (with the assumption that $k\neq 0$ everywhere in this section), we obtain a $USp(2\kappa)$ CS theory at level $-2N_c$ coupled to $2N_f$ squarks. The magnetic tree-level Lagrangian that was Higgsed leads automatically to a $USp(2N_f)$-invariant quartic interaction for the squarks of the form
\beq
\label{bosonab}
\LL_{quartic} = \left( \bar \phi^{ai} \phi_{aj} \right) \left( \bar \phi^{bj} \phi_{bi} \right) 
~.
\eeq 
This is one of the two possible quartic scalar interactions that are $USp(2N_f)$-invariant (see \cite{Aharony:2016jvv} for related comments; the interaction $\tilde \Omega^{ij} \phi_{aj} \Omega^{ab} \phi_{bk} \tilde \Omega^{kl} \phi_{cl} \Omega^{cd} \phi_{di}$ is the second possibility). The renormalization group will naturally induce this second interaction, as well as the $USp(2N_f)$-invariant mass term $\sum_{a,i} |\phi_{ai}|^2$. The match of massive deformations in the electric and magnetic theory above via suitable brane motions reinforces the picture that the bosonic theory is indeed at a Wilson-Fischer fixed point as required by bosonization. 

To summarise, after integrating out the gaugino, in the IR of the electric theory we have 
\beq
\label{bosonac}
USp(2N_c)_{2\kappa} ~\oplus ~ 2N_f ~{\rm quarks} 
~.
\eeq
In the IR of the magnetic theory we have
\beq
\label{bosonad}
USp(2\kappa)_{-2N_c} ~\oplus~ 2N_f ~ {\rm scalars}
~.
\eeq
In th regime of this subsection the non-supersymmetric Seiberg duality implies the 3d bosonization duality \eqref{introab} with the identification \eqref{introac}, $K=\kappa-\frac{N_f}{2}$.

\subsection{Symmetry breaking}
\label{partialcond}

Next we discuss the regions II and II$'$ in the phase diagram B in Fig.\ \ref{fig_phases}. We will argue in favour of a symmetry breaking scenario in both regions. Region II is characterised by the inequalities $N_f>\kappa>0$ and region II$'$ by the inequalities $-N_f <\kappa<0$. It is convenient to consider first region II$'$.

\subsubsection{Region II\,$'$} 

\begin{figure}[!t]
\centerline{\includegraphics[width=12cm]{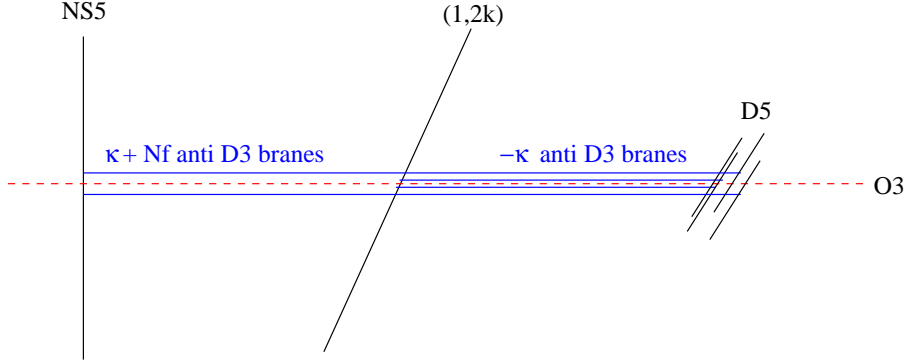}}
\caption{\footnotesize A realisation of the flavor breaking phase in the magnetic theory. $N_f+\kappa$ flavor branes reconnect with color branes. The magnetic gauge group is completely broken.}
\label{fig-breaking}
\end{figure}

$\kappa<0$ means that there are more flavor D3 branes than color D3 branes. As a result, at most $2N_f + 2\kappa$ D3 branes can reconnect. The color group is fully Higgsed and one is left with $2|\kappa|$ flavor D3s stretching between the $(1,2k)$ and the flavor D5s over an O3$^-$ plane in addition to the reconnected D3s, see Fig.\ref{fig-breaking}.

The elementary mesons on the $2|\kappa|$ D3s are massive. The flavor D3s stay attached to the orientifold and their global symmetry remains $SO(2|k|)$.

At the same time there is a global $SO(2(N_f + \kappa))$ symmetry associated with the $2(N_f+ \kappa)$ D3s.
We deduce that in string theory there is a breaking of the global $SO(2N_f)$ symmetry of the form
\beq
\label{pcondaa}
SO(2N_f) \to SO(2(N_f + \kappa)) \times SO(2|\kappa|) \, . 
\eeq

In the IR field theory the global symmetries enhance and it is not hard to show that the color-flavor locking vev of the squarks breaks the global $USp(2N_f)$ in the following way
\beq
\label{pcondac}
USp(2N_f) \to USp(2N_f+2\kappa) \times USp(2|\kappa|)
~.\eeq
This is exactly the same pattern of symmetry breaking 
\beq
\label{pcondad}
USp(2N_f) \to USp(N_f-2K) \times USp(N_f+2K)
\eeq
anticipated in \cite{Komargodski:2017keh} with the identification \eqref{introac}, $K=|\kappa|-\frac{N_f}{2}$. This identification is consistent with massive deformations of the theory.

Consequently, the IR physics of this phase is described by the coset $\sigma$-model of the Nambu-Goldstone modes associated with the breaking \eqref{pcondaa}. There are 
\bea 
& &
\tfrac{1}{2} 2N_f (2N_f-1) - \tfrac{1}{2} (2N_f +2\kappa)(2N_f + 2\kappa-1) -\tfrac{1}{2} 2|\kappa | (2|\kappa |-1) =  \nonumber  \\
& &=
4|\kappa | (N_f + \kappa) =   
(N_f-2K)(N_f+2K) \label{NG-bosons}
 \eea
massless Nambu-Goldstone bosons, which arise as massless modes on the open strings stretching between the $2|\kappa |$ D3-branes and the $2 (N_f + \kappa)$ D3-branes, see Fig.\ref{fig-breaking}. As a check, notice that the result \eqref{NG-bosons} (derived from \eqref{pcondaa}) agrees with the counting of Nambu-Goldstone bosons in QCD$_3$ \eqref{pcondad}. 
 
The order parameter for the breaking is $\langle \phi _{ai} \Omega ^{ab} \phi _{bj} \rangle$.

\subsubsection{Region II}

The region II, characterised by the parameter regime $N_f > \kappa>0$, is more intriguing. On one hand, this is a case where, at first sight, all $2N_f$ flavor D3s can reconnect with color D3s leading to an IR $USp(2\kappa)$ theory coupled to $2N_f$ bosons implying bosonization and no global symmetry breaking. On the other hand, the dictionary \eqref{introac} implies that this phase should still be describing the symmetry breaking phase of QCD$_3$ with $N_f > 2|K|$ where bosonization is inconsistent. It is hard to believe that the association of the orientifold QCD$_3$ with QCD$_3$ breaks down in this region because we can go to a limit of large $k$ where the gaugino becomes arbitrarily massive and can be integrated out safely. The only logical conclusion seems to be that something critical is happening to the brane configuration as one crosses the threshold $N_f= 2\kappa$. 

A natural guess is the presence of additional instabilities in the brane setup that describes the magnetic theory. Previously, in region I, it was sensible to expect that the brane reconnection leads to a stable vacuum. We would like to propose that the vacuum after brane reconnection is unstable, namely the $2N_f$ scalars in the resulting IR description have a negative mass squared. In that case, it seems likely that open string condensation could proceed by annihilating the $2\kappa$ color D3s against another set of $2\kappa$ color D3s eventually leaving behind $2(N_f-\kappa)$ reconnected D3s (between $2(N_f-\kappa)$ D5s and the NS5) and $2\kappa$ flavor D3s (between $2\kappa$ D5s and the $(1,2k)$ bound state). This is identical to the symmetry breaking configuration in region II$'$.

As a check of this scenario let us consider a deformation of the configuration with $N_f$ reconnected D3s and $\kappa$ D3s between the $(1,2k)$ bound state and the NS5 brane that adds mass to the scalars (equivalently, a real mass to the fermion quarks in the electric description). The deformation (see deformation 4 in Fig.\ \ref{massFig} in appendix \ref{appEvidence}) is implemented by bringing the $2N_f$ D5s on top of the $(1,2k)$ bound state and breaking them up to two half-D5s separated symmetrically in the $x^3$ direction. The $(1,2k)$ bound state connects with the broken D5s to create a $(1,2k+2)$ fivebrane bound state. The reconnected D3s are attached to the broken D5s and dragged along the $x^3$ direction. Hence, for a finite separation the low-lying scalar modes on the strings stretching between the color D3s and the reconnected D3s acquire a positive mass squared shift. If they are tachyonic at zero separation, as the above scenario dictates, then there is a critical distance where they become massless. Precisely at that point there is a dual description of the system in terms of the bosonic degrees of freedom. For larger separations the scalars are massive and can be integrated out to recover a topological QFT in the infra-red. This picture reproduces nicely all the features of the phase transition in Fig.\ 6 of Ref.\ \cite{Komargodski:2017keh}. Similar statements can be made for the mass deformation represented by the brane move 2 in Fig.\ \ref{massFig} in appendix \ref{appEvidence}.

The above-mentioned checks imply an overall picture which is consistent with the proposed scenario of a transition in brane physics at $N_f=\kappa$. It would be interesting to find further evidence in favour of this scenario with a more explicit computation in string theory.

\subsection{Scenario of symmetry restoration}
\label{massivesquark}

In QCD$_3$ it is expected \cite{Appelquist:1988sr,Appelquist:1989tc} that there is some $N_*(k,N_c)$ such that when $N_f\geq N_*$ the IR CFT is interacting and there is no symmetry breaking. Can this possibility arise naturally in our framework? We would like to argue that this could be described in the magnetic brane setup by a regime where the magnetic squarks are massive. 

Assume that quantum effects make the squarks massive (namely, their mass squared is positive).
Integrating out the massive squarks we obtain 
\beq
\label{highNfab}
\LL_{matter} =   i \bar\psi^{ai} \slashed{D} \psi_{ai} 
+ \p_\mu \bar M^{ij} \p^\mu M_{ij} + i \bar \chi^{ij} \slashed{\p} \chi_{ij}
- M^i_{~j} \psi_{ai} \bar\psi^{aj}
- \mu^2 \bar M^{ij} M_{ij} +\ldots
~,\eeq
where the dots indicate higher dimension interactions including an $M^4$ interaction. $\mu^2$ is an induced tachyonic mass squared to the mesons. To write this mass term we employed the proposed relation \eqref{mmrelation} between the signs of the mass squared of the squarks and the mesons, which suggests that the mesons are tachyonic in this case. Then, as the tachyonic mesons $M$ condense the quarks become massive. In the IR we obtain the sum of a topological QFT with free mesinos. In addition, at the true vacuum of the theory we have massive mesons, with $M^2 \sim 1/N_f$. Indeed, in the large-$N_f$ limit the theory on the flavor branes acquires supersymmetry because there is no distinction between the symmetric and the anti-symmetric representations. $1/N_f$ corrections should give a small mass to the mesons.

\section{The special case of Yang-Mills theory without a Chern-Simons term}
\label{yangmills}

Consider the special case where $k=0$. In the absence of a bare CS term the IR of the field theory is dominated by the Yang-Mills interaction. The theory is parity invariant classically. Due to the Vafa-Witten theorem \cite{Vafa:1984xg} that states that parity cannot be broken spontaneously, the theory is also parity invariant quantum mechanically. 

The brane dynamics is also expected to be different with respect to the case $k\ne 0$. Instead of having configurations with an NS5 brane and a tilted fivebrane we have configurations with parallel NS5 branes.

The magnetic theory is $USp(2N_f-2N_c+2)$. For $N_c \ge 1$, we have $N_f \ge N_f - N_c+1$, hence brane dynamics implies that the bosonization phase cannot occur. 

According to the philosophy advocated in this paper, whether flavour symmetry is broken or not depends on whether the magnetic squarks condense or become massive.

Brane dynamics suggests that if the squarks condense the flavour symmetry is broken $USp(2N_f)\rightarrow USp(2N_f-2N_c+2) \times USp(2N_c-2)$. This pattern, however, is not consistent with the Vafa-Witten theorem, because such a breaking pattern does not preserve parity.

According to eq.\ \eqref{squarkmass} the squarks may become tachyonic or massive. For large enough $k$ they become tachyonic. We propose that for $k=0$ the squarks are massive.

When the squarks become massive flavour symmetry is not broken. The dynamics is the same as that of the case $N_f \geq N_*$, which was described in section \eqref{massivesquark}. 
This result is different from the one for QCD$_3$ with $K=0$, as described in the literature \cite{Appelquist:1988sr,Appelquist:1989tc}. For sufficiently small $N_f$ QCD$_3$ without a Chern-Simons term is expected to break the flavour symmetry in a pattern consistent with parity $USp(2N_f)\rightarrow USp(N_f) \times USp(N_f)$. This disparity does not lead to an obvious inconsistency, because the present electric theory contains a {\it massless} gluino. The gluino does not acquire a CS mass and its presence can alter the IR dynamics. 

We therefore make a prediction, using brane dynamics, that 3d Yang-Mills theory with a fermion in the antisymmetric representation and $2N_f$ fundamental fermions does not break the USp flavour symmetry, as opposed to the theory without the antisymmetric fermion.

\subsection*{Acknowledgements}

We would like to thank Andreas Karch, Zohar Komargodski and David Tong for discussions. The work of AA has been supported by STFC grant ST/P00055X/1. The work of VN has been supported by STFC under the consolidated grant ST/P000371/1.

\begin{appendix}

\section{Evidence for non-supersymmetric Seiberg duality}
\label{appEvidence}

\subsection{'t Hooft anomaly matching}
\label{tHooft}

In this subsection we consider 't Hooft anomaly matching between the electric and magnetic sides of the orientifold QCD theory in section \ref{duality}. Both sides have a global $USp(2N_f)$ symmetry. We can gauge this symmetry by coupling to background $USp(2N_f)$ gauge fields. We want to identify the CS counterterms for these fields. We follow the steps outlined for $USp$ bosonization in \cite{Aharony:2016jvv} (related discussions about this type of anomalies can be found in \cite{Cherman:2017tey,Shimizu:2017asf,Gaiotto:2017tne}).

On the electric side we have $USp(2N_c)_{2k} \times USp(2N_f)_{2k_e}$.\footnote{In our notation $2k$ and $2k_e$ are the bare Chern-Simons levels.} On the magnetic side we have $USp(2\tilde N_c)\times USp(2N_f)_{2k_m}$. Both $k_e, k_m \in {\mathbb Z}$ and the second $USp$ factor is classical. Implementing the mass deformation 4 of Fig.\ \eqref{massFig} we get $USp(2N_c)_{2(k+ N_f)} \times USp(2N_f)_{2(k_e + N_c)}$ on the electric side. On the magnetic side we get $USp(2\tilde N_c)_{-2(k+ N_f)} \times USp(2N_f)_{2(k_m + \tilde N_c)}$. Integrating out the gaugino on the dynamical $USp(2N_c)$ and $USp(2\tilde N_c)$ factors we recover level-rank duality in the IR. The match of the CS levels of $USp(2N_f)$ on both sides requires
\beq
\label{anomaf}
k_e+N_c = k_m + \tilde N_c
~.
\eeq

The global symmetry that acts faithfully on local operators is $USp(2N_f)/{\mathbb Z}_2$. This puts restrictions on the CS counterterms. Consistency with the ${\mathbb Z}_2$ quotient on the electric side requires
\beq
\label{anomag}
N_c k + N_f k_e \in 2 {\mathbb Z}
~.
\eeq
On the magnetic side it requires 
\beq
\label{anomai}
- \tilde N_c k + N_f k_m \in 2 {\mathbb Z}
~.
\eeq
We notice that 
\bea
\label{anomaj}
- \tilde N_c k + N_f k_m 
&=& N_c k + N_f k_e + (N_c + \tilde N_c) (N_f -k) -2 \tilde N_c N_f
\nonumber\\
&=& N_c k + N_f k_e + (N_f + k + 1) (N_f -k) -2 \tilde N_c N_f
~.\eea
Since $(N_f + k + 1) (N_f -k) \in 2 {\mathbb Z}$ identically, we deduce that the conditions \eqref{anomag} and \eqref{anomai} are automatically satisfied together, which is a good consistency check.

\subsection{RG flows}
\label{RG}

\begin{figure}[!th]
\hspace{-1cm}
\begin{tikzpicture}
\draw (0,0);
\draw (1,2) node[circle,inner sep=1pt,draw] {1};
\draw[thick] (1,0) -- (5,0);
\draw[thick,dashed] (2,0) -- (2,1.5);
\draw[thick] (1.8,1.5) -- (2.2,1.5) -- (2.2,1.9) -- (1.8,1.9) -- (1.8,1.5);
\draw[thick,dashed] (2.9,0) -- (2.9,1.5);
\draw[thick,dashed] (3.1,0) -- (3.1,1.5) (3,2.3) node {\tiny $2(N_f-1)$} ;
\draw[thick] (2.8,1.5) -- (3.2,1.5) -- (3.2,1.9) -- (2.8,1.9) -- (2.8,1.5);
\draw[thick,red] (3,0) -- (3,3);
\draw[thick,dashed] (4,0) -- (4,1.5);
\draw[thick] (3.8,1.5) -- (4.2,1.5) -- (4.2,1.9) -- (3.8,1.9) -- (3.8,1.5);
\draw (2.9,0) -- (2.9,-1.5);
\draw (3.1,0) -- (3.1,-1.5);
\draw[thick,blue] (3,0) -- (3,-1.5) ;
\draw (3.5,-0.9) node {\tiny $2N_c$};
\draw (3,-1.7) circle (0.2cm) (3.7,-1.7) node {\tiny $(1,2k)$};
\draw [thick,red] (3,-1.9) -- (3,-2.5); 
\draw [thick] (1.4,-2.5) node {\it \small electric};
\draw[->] (5, 3) to (5,3.5) ;
\draw (5,3.8) node {\small 6};
\draw[->] (5,3) to (5.35,3);
\draw (5.7,3) node {\small 4,5};
\draw [thick] (5.9,-2.5) node {\it \small magnetic};
\draw[thick] (5.3,-1) -- (9.3,-1);
\draw[thick,red] (7.3,-1) -- (7.3,-2.5);
\draw (7.3,0.7) circle (0.2cm) (7.9,0.7) node {\tiny $(1,2k)$};
\draw[thick,blue] (7.3,-1) -- (7.3,0.5); 
\draw[thick] (7.2,-1) -- (7.2,0.5);
\draw[thick] (7.4,-1) -- (7.4,0.5);
\draw (8.1,-0.1) node {\tiny $2(\tilde N_c-1)$};
\draw[thick,dashed] (6,-1) -- (6,2);
\draw[thick] (5.8,2) -- (6.2,2) -- (6.2,2.4) -- (5.8,2.4) -- (5.8,2);
\draw[thick,dashed] (8.6,-1) -- (8.6,2);
\draw[thick] (8.4,2) -- (8.8,2) -- (8.8,2.4) -- (8.4,2.4) -- (8.4,2);
\draw[thick,red] (7.3,0.9) -- (7.3,3);
\draw[thick] (7.1,2) -- (7.5,2) -- (7.5,2.4) -- (7.1,2.4) -- (7.1,2);
\draw[thick,dashed] (7.2,0.9) -- (7.2,2);
\draw[thick,dashed] (7.4,0.9) -- (7.4,2);
\draw (8.1,1.4) node {\tiny $2(N_f-1)$};
\draw [thick] (9.9,-2.5) node {\it \small electric};
\draw[thick] (9.6,0) -- (13.3,0);
\draw[thick,dashed] (10.6,0) -- (10.6,1.5);
\draw[thick] (10.4,1.5) -- (10.8,1.5) -- (10.8,1.9) -- (10.4,1.9) -- (10.4,1.5);
\draw[thick,dashed] (11.4,0) -- (11.4,1.5);
\draw[thick,dashed] (11.6,0) -- (11.6,1.5) (11.7,2.3) node {\tiny $2(N_f-1)$} ;
\draw[thick] (11.3,1.5) -- (11.7,1.5) -- (11.7,1.9) -- (11.3,1.9) -- (11.3,1.5);
\draw[thick,red] (11.5,0) -- (11.5,3);
\draw[thick,dashed] (12.3,0) -- (12.3,1.5);
\draw[thick] (12.1,1.5) -- (12.5,1.5) -- (12.5,1.9) -- (12.1,1.9) -- (12.1,1.5);
\draw (11.4,0) -- (11.4,-1.5);
\draw (11.6,0) -- (11.6,-1.5);
\draw[thick,blue] (11.5,0) -- (11.5,-1.5) ;
\draw (12,-0.9) node {\tiny $2N_c$};
\draw [thick] (9.6,-1.5) -- (13.3,-1.5);
\draw [thick] (9.6,-1.9) -- (13.3,-1.9);
\draw (11.5,-1.7) node {\tiny $(1,2k)$};
\filldraw[fill=gray,opacity=0.1] (9.6,-1.9) rectangle (13.3,-1.5);
\draw [thick,red] (11.5,-1.9) -- (11.5,-2.5); 
\draw [thick] (13.7,-2.5) node {\it \small magnetic};
\draw (17.2,2) node[circle,inner sep=1pt,draw] {2};
\draw[thick] (13.4,-1) -- (17.1,-1);
\draw[thick,red] (15.2,-1) -- (15.2,-2.5);
\draw [thick] (13.4,0.5) -- (17.1,0.5);
\draw [thick] (13.4,0.9) -- (17.1,0.9);
\draw (15.2,0.7) node {\tiny $(1,2k)$};
\filldraw[fill=gray,opacity=0.1] (13.4,0.5) rectangle (17.1,0.9);
\draw[thick,blue] (15.2,-1) -- (15.2,0.5); 
\draw[thick] (15.1,-1) -- (15.1,0.5);
\draw[thick] (15.3,-1) -- (15.3,0.5);
\draw (15.9,-0.1) node {\tiny $2(\tilde N_c-1)$};
\draw[thick,dashed] (13.9,-1) -- (13.9,2);
\draw[thick] (13.7,2) -- (14.1,2) -- (14.1,2.4) -- (13.7,2.4) -- (13.7,2);
\draw[thick,dashed] (16.5,-1) -- (16.5,2);
\draw[thick] (16.3,2) -- (16.7,2) -- (16.7,2.4) -- (16.3,2.4) -- (16.3,2);
\draw[thick,red] (15.2,0.9) -- (15.2,3);
\draw[thick] (15.0,2) -- (15.4,2) -- (15.4,2.4) -- (15.0,2.4) -- (15.0,2);
\draw[thick,dashed] (15.1,0.9) -- (15.1,2);
\draw[thick,dashed] (15.3,0.9) -- (15.3,2);
\draw (16.0,1.4) node {\tiny $2(N_f-1)$};
\draw[->] (13.2, 3) to (13.2,3.5) ;
\draw (13.2,3.8) node {\small 6};
\draw[->] (13.2,3) to (13.55,3);
\draw (13.7,3) node {\small 3};
\end{tikzpicture}

\vspace{1cm}
\hspace{-1cm}
\begin{tikzpicture}
\draw (0,0);
\draw (1,2) node[circle,inner sep=1pt,draw] {3};
\draw[thick] (1,0) -- (5,0);
\draw[thick,dashed] (2,-1.5) -- (2,1.5);
\draw[thick] (1.8,1.5) -- (2.2,1.5) -- (2.2,1.9) -- (1.8,1.9) -- (1.8,1.5);
\draw[thick,dashed] (2.9,0) -- (2.9,1.5);
\draw[thick,dashed] (3.1,0) -- (3.1,1.5) (3,2.3) node {\tiny $2(N_f-1)$} ;
\draw[thick] (2.8,1.5) -- (3.2,1.5) -- (3.2,1.9) -- (2.8,1.9) -- (2.8,1.5);
\draw[thick,red] (3,0) -- (3,3);
\draw[thick,dashed] (4,-1.5) -- (4,1.5);
\draw[thick] (3.8,1.5) -- (4.2,1.5) -- (4.2,1.9) -- (3.8,1.9) -- (3.8,1.5);
\draw (2.9,0) -- (2.9,-1.5);
\draw (3.1,0) -- (3.1,-1.5);
\draw[thick,blue] (3,0) -- (3,-1.5) ;
\draw (3.8,-0.9) node {\tiny $2(N_c-1)$};
\draw [thick] (1,-1.5) -- (5,-1.5);
\draw [thick] (1,-1.9) -- (5,-1.9);
\draw (3,-1.7) node {\tiny $(1,2k)$};
\filldraw[fill=gray,opacity=0.1] (1,-1.9) rectangle (5,-1.5);
\draw [thick,red] (3,-1.9) -- (3,-2.5); 
\draw [thick] (1.4,-2.5) node {\it \small electric};
\draw[->] (5, 3) to (5,3.5) ;
\draw (5,3.8) node {\small 6};
\draw[->] (5,3) to (5.35,3);
\draw (5.6,3) node {\small 3};
\draw [thick] (5.9,-2.5) node {\it \small magnetic};
\draw[thick] (5.3,-1) -- (9.3,-1);
\draw[thick,red] (7.3,-1) -- (7.3,-2.5);
\draw [thick] (5.3,0.5) -- (9.3,0.5);
\draw [thick] (5.3,0.9) -- (9.3,0.9);
\draw (7.3,0.7) node {\tiny $(1,2k)$};
\filldraw[fill=gray,opacity=0.1] (5.3,0.5) rectangle (9.3,0.9);
\draw[thick,blue] (7.3,-1) -- (7.3,0.5); 
\draw[thick] (7.2,-1) -- (7.2,0.5);
\draw[thick] (7.4,-1) -- (7.4,0.5);
\draw (7.8,-0.1) node {\tiny $2\tilde N_c$};
\draw[thick,dashed] (6,0.9) -- (6,2);
\draw[thick] (5.8,2) -- (6.2,2) -- (6.2,2.4) -- (5.8,2.4) -- (5.8,2);
\draw[thick,dashed] (8.6,0.9) -- (8.6,2);
\draw[thick] (8.4,2) -- (8.8,2) -- (8.8,2.4) -- (8.4,2.4) -- (8.4,2);
\draw[thick,red] (7.3,0.9) -- (7.3,3);
\draw[thick] (7.1,2) -- (7.5,2) -- (7.5,2.4) -- (7.1,2.4) -- (7.1,2);
\draw[thick,dashed] (7.2,0.9) -- (7.2,2);
\draw[thick,dashed] (7.4,0.9) -- (7.4,2);
\draw (8.1,1.4) node {\tiny $2(N_f-1)$};
\draw [thick] (9.9,-2.5) node {\it \small electric};
\draw[thick] (9.6,0) -- (13.3,0);
\draw[thick,dashed] (11.4,0) -- (11.4,1.5);
\draw[thick,dashed] (11.6,0) -- (11.6,1.5) (11.7,2.3) node {\tiny $2(N_f-1)$} ;
\draw[thick] (11.3,1.5) -- (11.7,1.5) -- (11.7,1.9) -- (11.3,1.9) -- (11.3,1.5);
\draw[thick,red] (11.5,0) -- (11.5,3);
\draw (11.4,0) -- (11.4,-1.5);
\draw (11.6,0) -- (11.6,-1.5);
\draw[thick,blue] (11.5,0) -- (11.5,-1.5) ;
\draw (12,-0.9) node {\tiny $2N_c$};
\draw (11.5,-1.7) circle (0.2cm) (12.4,-1.7) node {\tiny $(1,2k+2)$};
\draw [thick,red] (11.5,-1.9) -- (11.5,-2.5); 
\draw [thick] (13.7,-2.5) node {\it \small magnetic};
\draw (17.2,2) node[circle,inner sep=1pt,draw] {4};
\draw[thick] (13.4,-1) -- (17.1,-1);
\draw[thick,red] (15.2,-1) -- (15.2,-2.5);
\draw (15.2,0.7) circle (0.2cm) (16.1,0.7) node {\tiny $(1,2k+2)$};
\draw[thick,blue] (15.2,-1) -- (15.2,0.5); 
\draw[thick] (15.1,-1) -- (15.1,0.5);
\draw[thick] (15.3,-1) -- (15.3,0.5);
\draw (15.9,-0.1) node {\tiny $2\tilde N_c$};
\draw[thick,red] (15.2,0.9) -- (15.2,3);
\draw[thick] (15.0,2) -- (15.4,2) -- (15.4,2.4) -- (15.0,2.4) -- (15.0,2);
\draw[thick,dashed] (15.1,0.9) -- (15.1,2);
\draw[thick,dashed] (15.3,0.9) -- (15.3,2);
\draw (16.0,1.4) node {\tiny $2(N_f-1)$};
\draw[->] (13.2, 3) to (13.2,3.5) ;
\draw (13.2,3.8) node {\small 6};
\end{tikzpicture}

	\caption[mass]
	{\footnotesize Summary of mass deformations in the electric and magnetic brane setups. The red lines denote the O3$^-$ planes and the blue lines the O3$^+$ planes. The horizontal solid black line in the middle of each diagram is the NS5 brane. The circle in diagrams 1 and 4, and the grey belt in diagrams 2 and 3 is the fivebrane bound state. The vertical solid black lines are the color D3 branes and the vertical dashed black lines are the flavor D3 branes. The vertical direction is always along $x^6$ and the horizontal along the (45) plane in diagram 1 and along $x^3$ in diagrams 2 and 3.}
	\label{massFig}
\end{figure}
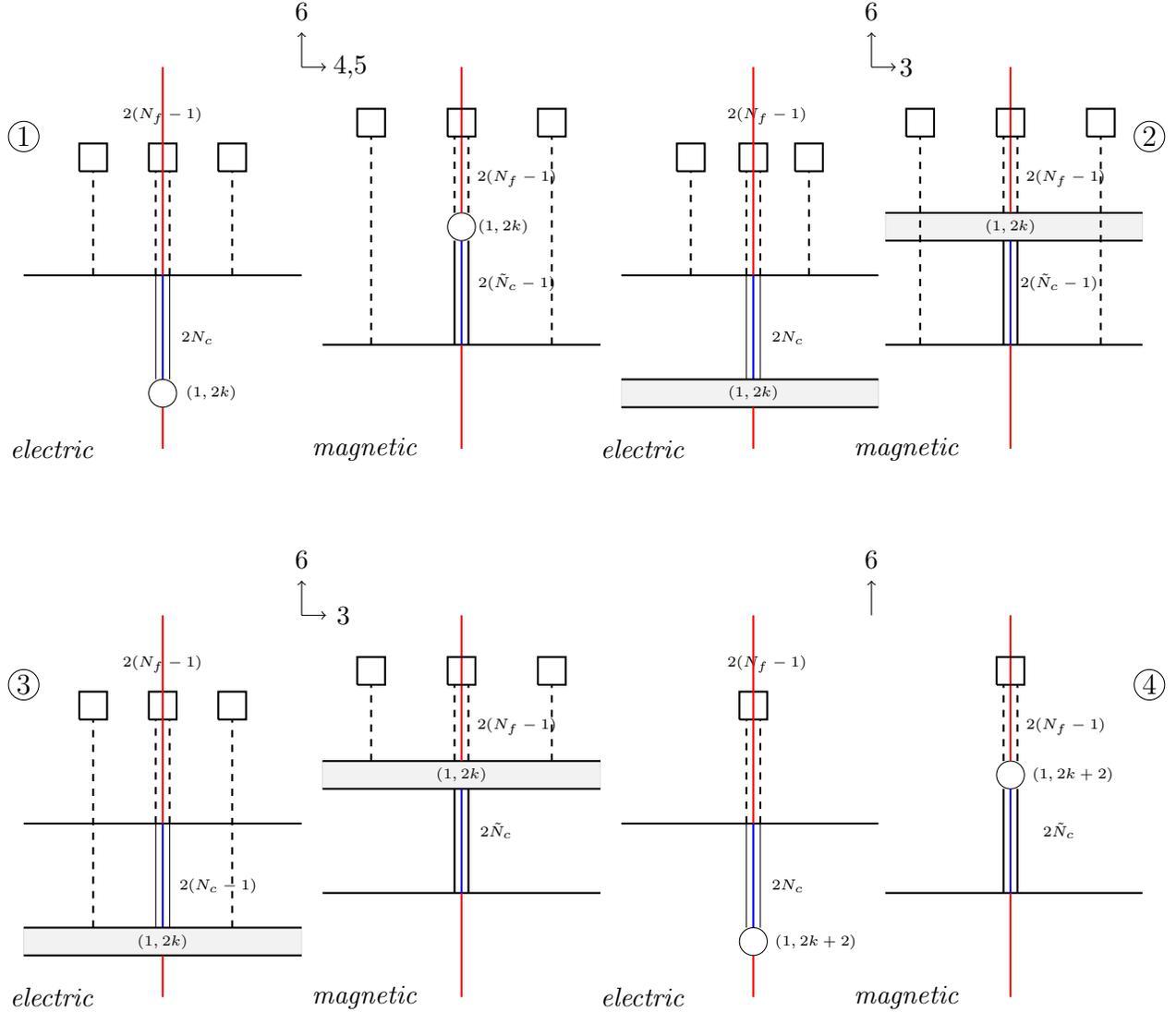

In the tree-level orientifold QCD$_3$ theory we can consider four types of mass deformations. In this appendix we mostly ignore the potential quantum-induced instabilities. When such instabilities exist we implement the deformations at the unstable vacuum. It is convenient to see first how the deformations of interest map between the electric and magnetic descriptions in the brane setup. See \cite{Giveon:2008zn} for analogous deformations in a $d=3$ $\NN=2$ setup. The brane deformations are summarised pictorially in Fig.\ \ref{massFig}.

The first deformation, number 1 in Fig.\ \ref{massFig}, moves an image pair of D5s in the (45) plane. On the electric side this reduces $N_f \to N_f-1$ while $N_c \to N_c$ and $k \to k$. On the field theory side this involves a complex mass deformation of the form
\beq
\label{rgaa}
\delta \LL_{electric} = 
m^{11} \bar m_{11} \bar\phi^{a1} \phi_{a1}
- m^{11} \psi^a_1 \psi_{a1} - \bar m_{11} \, \bar \psi^{a1} \bar \psi_a^{~1}
~,
\eeq
which indeed removes one flavor, $N_f \to N_f -1$, but does not shift the level $k$ or the rank of the gauge group $N_c$.
On the magnetic side the deformation has the effect $N_f \to N_f-1$, $\tilde N_c \to \tilde N_c -1$, $k \to k$. The corresponding deformation gives a vev to $\phi_{a1}$ so that
\beq
\label{rgac}
\phi_{a1}\phi^a_{~1} = - \bar m_{11}
~.\eeq
The vev of $\phi_{a1}$ breaks the magnetic gauge group, $\tilde N_c \to \tilde N_c -1$, removes one flavor ($N_f \to N_f-1$) and leaves $k$ invariant as anticipated.

The second deformation, number 2 in Fig.\ \ref{massFig}, moves an image pair of D5s in the 3 direction dragging along a pair of flavor D3s stretching between a D5 and the NS5. On the electric side this reduces $N_f \to N_f -1$ while $N_c \to N_c$ and $k \to k$. On the field theory side this involves equal and opposite sign masses for one flavor
\bea
\label{rgac2}
\delta \LL_{electric} = 
m\, \bar\psi^{a, N_{f+1}} \psi_{a1} + m\, \bar\psi^{a1} \psi_{a,N_f+1} 
= m\, \psi^a_{~1} \psi_{a1} - m \, \psi^{a,N_f+1} \psi_{a,N_f+1} 
~.
\eea
The scalars are not involved in this deformation. Indeed, \eqref{rgac2} removes the fermions of one flavor and after integrating out all auxiliary fields we find a potential for the scalars $\phi_{a1}$ with a quartic and sextic term. There are two potential vacua for this potential. One at the origin and one away from the origin. The deformation number 2 in Fig.\ \ref{massFig} does not Higgs the gauge group so it corresponds to the vacuum at the origin. 
On the magnetic side the moved flavor D3s reconnect with 2 image color D3s and stretch between the D5s and the NS5. This reduces $N_f \to N_f-1$, $\tilde N_c \to \tilde N_c-1$ and keeps $k\to k$. 

The third deformation, number 3 in Fig.\ \ref{massFig}, moves an image pair of D5s in the 3 direction but now in the electric side the corresponding flavor D3s reconnect with the $(1,2k)$ 5 brane. This deformation reduces $N_f \to N_f-1$, $N_c \to N_c -1$ and keeps $k\to k$. On the field theory side this is still a case of equal and opposite sign masses for one flavor, see eq.\ \eqref{rgac}. In this case the scalars get a vev and the gauge group is Higgsed, i.e. $N_c \to N_c-1$. On the magnetic side the moved flavor D3s stretch between the D5s and the $(1,2k)$ bound state. This reduces $N_f \to N_f-1$ but leaves $\tilde N_c$ and $k$ intact. Notice that this move is only possible if the flavor and color D3s are not reconnected. If the magnetic squarks are tachyonic, the move would have to be implemented on an unstable vacuum. 

The final deformation, number 4 in Fig.\ \ref{massFig}, moves a pair of flavor D5s along the 6 direction on the $(1,2k)$ bound state where they merge to form a $(1,2k+2)$ bound state. 
This operation reduces $N_f \to N_f -1$ but $N_c \to N_c$ and $k \to k+ 1$. On the field theory side this operation involves giving same sign real masses 
to one flavor 
\beq
\label{rgba}
\delta \LL_{electric} = m \, \bar \psi^{a1} \psi_{a1} +m\, \bar \psi^{a, N_f+1} \psi_{a, N_f+1}
= 2 m \, \bar \psi^{a1} \psi_{a1}
~.
\eeq
On the magnetic side the magnetic quarks acquire corresponding masses and as these masses are sent to infinity $N_f \to N_f -1$, $\tilde N_c \to \tilde N_c$ and $k \to k+ 1$ consistently with the duality.

\section{Unitary groups}
\label{unitary}

\begin{table}[!t]
\begin{center}
\begin{tabular}{|c|c|c|}
\hline
\multicolumn{3}{|c|} {Electric Theory} \\ 
\hline \hline
 & $U(2N_c)$ & $U(2N_f)$  \\
\hline
 $A_\mu$ & adjoint & \bdot \\
& $(2N_c)^2\B$ & \\
\hline
 $\sigma $ & adjoint & \bdot \\
& $(2N_c)^2\B$ & \\
\hline
$\lambda$ & $\T\Yasymm\B$   &  \bdot  \\
& $2N_c(2N_c-1)\B$ & \\
\hline
$\Phi$   & $\Yfund$ & $\Yfund$ \\
&   $2N_c$  &   $2N_f$ \\
\hline
$\Psi$ & $\Yfund$ & $\Yfund$ \\
&   $2N_c$  &   $2N_f$  \\                   
\hline
\end{tabular}
\caption{\footnotesize The matter content of the electric theory. }
\label{Utableelectric}
\end{center}
\end{table}

In this section we propose a duality for a $U(2N_c)$ orientifold QCD$_3$ theory. In order to have a string theory realisation of a non-supersymmetric theory and a Seiberg duality, we follow \cite{Armoni:2014cia} where the $N_f=0$ case was considered (see also \cite{Armoni:2008gg,Armoni:2013kpa} for the analogous 4d theory).

The field theory lives on a type 0B brane configuration. The electric and magnetic brane configurations are identical to those we considered in section \ref{duality}. The orientifold projection is special to type 0 strings: it decouples the bulk tachyon from the brane and projects out half of the RR fields\cite{Sagnotti:1995ga,Sagnotti:1996qj}.\footnote{A more careful treatment of this setup requires its formulation in a {\it non-critical} type 0 string theory with Sagnotti-type orientifolds. This construction was performed in the context of the analogous 4d theory in \cite{Armoni:2008gg} following earlier work on non-supersymmetric non-critical strings in \cite{Israel:2007nj}. In that construction the non-critical type 0 string theory is classically stable and is not plagued by closed string tachyons. The corresponding formulation of a non-critical string theory for the 3d theory at hand would require exact worldsheet techniques for backgrounds with RR fluxes, which are currently beyond reach.} The matter content of the electric theory on the brane configuration is given in table \ref{Utableelectric}. In the planar (large-$N$) limit the theory is equivalent to the corresponding ${\cal N}=2$ $U(2N)$ gauge theory that lives on type IIB branes. Note also that the $U(2N_f)$ global symmetry is the symmetry of the IR theory. 

By swapping the fivebranes we arrive at the magnetic theory. Its matter content is given in table \ref{Utablemagnetic}.

\begin{table}[!t]
\begin{center}
\begin{tabular}{|c|c|c|}
\hline
\multicolumn{3}{|c|} {Magnetic Theory} \\ 
\hline \hline
 & $\T U(2\tilde N_c)$ & $U(2N_f)$  \\
\hline
 $a_\mu$ & adjoint & \bdot \\
& $ (2\tilde N_c)^2\B$ & \\
\hline
 $s$ & adjoint & \bdot \\
& $(2\tilde N_c)^2\B$ & \\
\hline
$l$ & $\T\Yasymm\B$   &  \bdot  \\
& $2\tilde N_c(2\tilde N_c-1)\B$ & \\
                   \hline
$\phi$   & $\Yfund$ & ${\Yfund}$ \\
&   $2\tilde N_c$  &   $2{N_f}$ \\
\hline
$\psi$ & $\Yfund$ & ${\Yfund}$ \\
&   $2\tilde N_c$  &   $2{N_f}$  \\ 
         \hline
 $M$ & \bdot & \Yasymm \\
& & $2N_f(2N_f-1)$  \\
\hline
$\chi$ & \bdot &  adjoint    \\
& & $(2N_f)^2$ \\         
\hline
\end{tabular}
\caption{\footnotesize The matter content of the magnetic theory. $\tilde N_c=N_f+k-N_c+1$.}
\label{Utablemagnetic}
\end{center}
\end{table}

The dynamics of the theory based on unitary groups is similar to the dynamics of the theory based on USp groups. This is not surprising: there exists non-perturbative planar equivalence between $USp(2N)$ and $U(2N)$ gauge theories.

On the electric side the scalars (squark and scalar gaugino) acquire mass and decouple. The CS level is shifted due to the antisymmetric gluino.

The IR levels of the  electric theory $2K_{\rm SU}$, $2K_{\rm U(1)}$ are given by 
\beq 
 K_{\rm SU} = |k+1-N_c| - \frac{N_f}{2} \,\,\, ; \,\,\, K_{\rm U(1)} = k+1 - \frac{N_f}{2} \,
 \eeq
 thanks to the fundamental fermions and the antisymmetric fermion that shift the bare level $2k$.  Note that in the bosonized phase $K_{\rm U(1)} = K_{\rm SU} + N_c$. 
 
In the preliminary discussion of this appendix we single out the case with even color, $2N_c$ and even flavor, $2N_f$, which proceeds as in the USp case.
 
When $N_f \le 2K_{\rm SU}$ the squarks condense, the Higgs mechanism takes place, and we find a dual bosonized theory of the form $U(2K_{\rm SU}+ N_f)$ with shifted levels 
 \beq 
 \tilde K_{\rm SU} =  -N_c \,\,\, ; \,\,\, \tilde K_{\rm U(1)} = -N_c + K_{\rm SU} +\frac{N_f}{2} \, .
 \eeq
  When $ 2K_{\rm SU} \le N_f \le N_{*}$ the reconnected branes suggest that
  $U(2N_f) \rightarrow U(N_f +2 K_{\rm SU}) \times U(N_f- 2K_{\rm SU} )$, in agreement with the conjecture by KS \cite{Komargodski:2017keh}. There are 
\bea 
& &
 ( 2N_f)^2 -  (2N_f +2\kappa)^2 - (2\kappa)^2  =  \nonumber  \\
 & &= 8|\kappa | (N_f + \kappa) =   
2(N_f-2K_{\rm SU})(N_f+2K_{\rm SU}) \label{NG-bosons2}
 \eea
massless Nambu-Goldstone bosons that corresponds to massless modes of the oriented open strings stretched between the $2|\kappa |$ branes and $2 (N_f + \kappa)$ branes

  The case $k=0$ (or $K_{\rm SU}$=0) is special. As in the USp theory we propose that there is no symmetry breaking and the theory is described by the magnetic theory of $N_f\geq N_*$. For $N_f \geq N_*$ the magnetic theory consists of a massless mesino and a light meson of mass square $1/N_f$.

\end{appendix}

\newpage
\providecommand{\href}[2]{#2}

\bibliographystyle{utphys}

\end{document}